\documentclass[traditabstract,longauth]{aa} 

\usepackage{graphicx}
\usepackage{txfonts}
\usepackage{longtable}
\begin{document}

   \title{A Virgo Environmental Survey Tracing Ionised Gas Emission (VESTIGE).I. Introduction to the Survey\thanks{Based on observations obtained with
   MegaPrime/MegaCam, a joint project of CFHT and CEA/DAPNIA, at the Canadian-French-Hawaii Telescope
   (CFHT) which is operated by the National Research Council (NRC) of Canada, the Institut National
   des Sciences de l'Univers of the Centre National de la Recherche Scientifique (CNRS) of France and
   the University of Hawaii.}
      }
   \subtitle{}
  \author{A. Boselli\inst{1},  
          M. Fossati\inst{2,3},  
          L. Ferrarese\inst{4},
	  S. Boissier\inst{1},
	  G. Consolandi\inst{5,6}, 
	  A. Longobardi\inst{7},
	  P. Amram\inst{1},
	  M. Balogh\inst{8},
	  P. Barmby\inst{9},
	  M. Boquien\inst{10},
	  F. Boulanger\inst{11},
	  J. Braine\inst{12},
	  V. Buat\inst{1},
	  D. Burgarella\inst{1},
	  F. Combes\inst{13,14},         
	  T. Contini\inst{15},         
          L. Cortese\inst{16},   
          P. C{\^o}t{\'e}\inst{4},
	  S. C{\^o}t{\'e}\inst{4},
          J.C. Cuillandre\inst{17},
	  L. Drissen\inst{18},
	  B. Epinat\inst{1},
          M. Fumagalli\inst{19}, 
	  S. Gallagher\inst{8}, 
          G. Gavazzi\inst{5},
	  J. Gomez-Lopez\inst{1},
          S. Gwyn\inst{4},   
	  W. Harris\inst{20},   
          G. Hensler\inst{21},
	  B. Koribalski\inst{22},
	  M. Marcelin\inst{1},
	  A. McConnachie\inst{4},
	  M.A. Miville-Deschenes\inst{11},
	  J. Navarro\inst{23},
	  D. Patton\inst{24},
	  E.W. Peng\inst{7,25},
	  H. Plana\inst{26},
	  N. Prantzos\inst{27},
	  C. Robert\inst{18},
	  J. Roediger\inst{4},
	  Y. Roehlly\inst{28},
	  D. Russeil\inst{1},
	  P. Salome\inst{12},
	  R. Sanchez-Janssen\inst{29},
	  P. Serra\inst{30},
	  K. Spekkens\inst{31,32},
          M. Sun\inst{33},
	  J. Taylor\inst{8},
          S. Tonnesen\inst{34},
	  B. Vollmer\inst{35},
	  J. Willis\inst{23},
	  H. Wozniak\inst{36}, 
	  T. Burdullis\inst{37}, 
	  D. Devost\inst{37}, 
	  B. Mahoney\inst{37}, 
	  N. Manset\inst{37}, 
	  A. Petric\inst{37}, 
	  S. Prunet\inst{37}, 
	  K. Withington.\inst{37}
       }

\institute{     
                Aix Marseille Univ, CNRS, LAM, Laboratoire d'Astrophysique de Marseille, Marseille, France
                \email{alessandro.boselli@lam.fr}
        \and
                Max-Planck-Institut f\"{u}r Extraterrestrische Physik, Giessenbachstrasse, 85748, Garching, Germany 
                \email{mfossati@mpe.mpg.de}
        \and  
                Universit{\"a}ts-Sternwarte M{\"u}nchen, Scheinerstrasse 1, D-81679 M{\"u}nchen, Germany
        \and
                NRC Herzberg Astronomy and Astrophysics, 5071 West Saanich Road, Victoria, BC, V9E 2E7, Canada
                \email{laura.ferrarese@nrc-cnrc.gr.ca}
	\and
                Universit\'a di Milano-Bicocca, piazza della scienza 3, 20100, Milano, Italy
        \and
		INAF - Osservatorio Astronomico di Brera, via Brera 28, 20159 Milano, Italy
	\and
		Kavli Institute for Astronomy and Astrophysics, Peking University, Beijing 100871, China 
	\and
		Department of Physics and Astronomy, University of Waterloo, Waterloo, Ontario N2L 3G1, Canada
	\and
		Department of Physics \& Astronomy, University of Western Ontario, London, ON N6A 3K7, Canada
	\and
		Unidad de Astronom\'ia, Fac. de Ciencias Basicas, Universidad de Antofagasta, Avda. U. de Antofagasta 02800, Antofagasta, Chile
	\and
		Institut d'Astrophysique Spatiale, UMR 8617, Universit\'e Paris-Sud, Batiment 121, 91405, Orsay, France
	\and
		Laboratoire d'Astrophysique de Bordeaux, Univ. Bordeaux, CNRS, B18N, all\'ee Geoffroy Saint-Hilaire, 33615 Pessac, France
	\and
		College de France, 11 Pl. M. Berthelot, F-75005 Paris, France 
	\and
		LERMA, Observatoire de Paris, CNRS, PSL Research University, Sorbonne Universit\'es, UPMC Univ. Paris 06, F-75014 Paris, France
        \and    
		IRAP, Institut de Recherche en Astrophysique et Plan\'tologie, CNRS, 14 avenue Edouard Belin, 31400, Toulouse, France
	\and
                International Centre for Radio Astronomy Research, The University of Western Australia, 35 Stirling Highway, Crawley WA 6009, Australia
        \and        
		CEA/IRFU/SAP, Laboratoire AIM Paris-Saclay, CNRS/INSU, Universit\'e Paris Diderot, Observatoire de Paris, PSL Research University, F-91191 Gif-sur-Yvette Cedex, France
        \and	
		Universit\'e Laval, 2325, rue de l'universit\'e, Qu\'ebec (Qu\'ebec), G1V 0A6, Canada
        \and
                Institute for Computational Cosmology and Centre for Extragalactic Astronomy, Department of Physics, Durham University, South Road, Durham DH1 3LE, UK
	\and
		Department of Physics \& Astronomy, McMaster University, Hamilton, ON, Canada
	\and
                Department of Astrophysics, University of Vienna, T\"urkenschanzstrasse 17, 1180, Vienna, Austria
	\and
		Australia Telescope National Facility, CSIRO Astronomy and Space Science, P.O. Box 76, Epping, NSW 1710
	\and
		Department of Physics and Astronomy, University of Victoria, PO Box 1700 STN CSC, Victoria, BC V8W 2Y2, Canada
	\and
		Department of Physics and Astronomy, Trent University, 1600 West Bank Drive, Peterborough, ON K9L 0G2, Canada
	\and
		Department of Astronomy, Peking University, Beijing 100871, China 
	\and
		Laboratorio de Astrofisica Teorica e Observacional, Universidade Estadual de Santa Cruz - 45650-000, Ilh\'eus-BA, Brasil
	\and
		Institut d'Astrophysique de Paris, UMR7095 CNRS, Universit\'e P. \& M. Curie, 98bis Bd. Arago, F-75104 Paris, France
	\and
		Astronomy Centre, Department of Physics and Astronomy, University of Sussex, Falmer, Brighton BN1 9QH, UK
	\and
		UK Astronomy Technology Centre, Royal Observatory Edinburgh, Blackford Hill, Edinburgh, EH9 3HJ, UK
	\and
		Osservatorio Astronomico di Cagliari, via della scienza 5, 09047 Selargius, Cagliari, Italy
	\and
		Department of Physics, Royal Military College of Canada, P.O. Box 17000, Station Forces, Kingston, ON K7L 7B4, Canada
	\and	
		Department of Physics, Engineering Physics, and Astronomy, Queen's University, Kingston, ON K7L 3N6, Canada 
	\and
                Department of Physics and Astronomy, University of Alabama in Huntsville, Huntsville, AL 35899, USA
	\and
		Center for Computational Astrophysics, Flatiron Institute, 162 5th Avenue, New York, NY 10003, USA
	\and
		Observatoire Astronomique de Strasbourg, UMR 7750, 11, rue de l'Universit\'e, 67000, Strasbourg, France
	\and
		LUPM, Univ. Montpellier, CNRS, Montpellier, France
	\and
		Canada-Frence-Hawaii Telescope Corporation, Kamuela, HI96743, USA
                   }

\authorrunning{Boselli et al.}
\titlerunning{VESTIGE}

   \date{}

 
  \abstract  
{The Virgo Environmental Survey Tracing Ionised Gas Emission (VESTIGE) is a blind narrow-band H$\alpha$+[NII]
imaging survey carried out with MegaCam at the Canada-France-Hawaii Telescope. The survey covers the whole Virgo 
cluster region from its core to one virial radius (104 deg$^2$). The sensitivity of the survey is of
$f(H\alpha)$ $\sim$ 4 $\times$ 10$^{-17}$ erg sec$^{-1}$ cm$^{-2}$ (5$\sigma$ detection limit) for point sources and
$\Sigma (H\alpha)$ $\sim$ 2 $\times$ 10$^{-18}$ erg sec$^{-1}$ cm$^{-2}$ arcsec$^{-2}$ (1$\sigma$ detection limit at 3 arcsec resolution) 
for extended sources, making VESTIGE the deepest and largest blind narrow-band survey of a nearby cluster.
This paper presents the survey in all its technical aspects, including the survey design, the observing strategy, 
the achieved sensitivity in both the narrow-band H$\alpha$+[NII] and in the broad-band $r$ filter used for the
stellar continuum subtraction, the data reduction, calibration, and products, as well as its status after the first 
observing semester. We briefly describe the H$\alpha$ properties of galaxies located in a 4$\times$1 deg$^2$ strip 
in the core of the cluster north of M87, where several extended tails of ionised gas are detected. This paper also lists 
the main scientific motivations of VESTIGE, which include the study of the effects of the environment on galaxy
evolution, the fate of the stripped gas in cluster objects, the star formation process in nearby galaxies of
different type and stellar mass, the determination of the H$\alpha$ luminosity function and of the
H$\alpha$ scaling relations down to $\sim$ 10$^6$ M$_{\odot}$ stellar mass objects, and the reconstruction 
of the dynamical structure of the Virgo cluster. This unique set of data will also be used to study the HII
luminosity function in hundreds of galaxies, the diffuse H$\alpha$+[NII] emission of 
the Milky Way at high Galactic latitude, and the properties of emission line galaxies at high redshift.}
   {}
   {}
   {}
   {}
   {}


   \maketitle
%

\section{Introduction}

Understanding the formation and evolution of galaxies remains a primary goal of modern astrophysics. The study of large
samples of galaxies detected in wide field, multifrequency, ground- and space-based surveys, both in the local Universe 
(SDSS - York et al. 2000, GALEX - Martin et al. 2005, 2MASS - Skrutskie et al. 2006, ALFALFA - Giovanelli et al. 2005, 
HIPASS - Meyer et al. 2004, NVSS - Condon et al. 1998, WISE - Wright et al. 2010, all sky surveys) and at high redshift, 
has led to significant progress towards an understanding of the process of galaxy evolution. The 
sensitivity, angular and spectral resolutions of the multifrequency data obtained in the most recent surveys have been 
fundamental in tracing the physical properties of different galaxy components, e.g., stellar populations, gas in its 
different phases (cold atomic and molecular, ionised, hot), heavy elements (metals and dust), and dark matter, whose 
content and distribution are tightly connected to the evolutionary state of galaxies (e.g., Boselli 2011). These achievements have been 
mirrored by advances in the speed and precision of numerical methods used to simulate the formation of structures over 
wide ranges in mass and radius (e.g. Vogelsberger et al. 2014; Genel et al. 2014; Schaye et al. 2015; Crain et al. 2015).    

Both observations and simulations consistently point to two main factors as key 
drivers of galaxy evolution: the secular evolution mainly driven by the dynamical mass of the system (e.g. Cowie et al. 1996; 
Gavazzi et al. 1996; Boselli et al. 2001) and the environment in which galaxies reside (Dressler 1980; Dressler et al. 1997; Balogh et al. 2000;  
Kauffmann et al. 2004; Boselli \& Gavazzi 2006, 2014; Peng et al. 2010). 
The relative importance of these two factors over cosmic timescales for systems of different mass and type, however, 
remains elusive. Further progress hinges on the characterisation of astrophysical processes that are not fully understood 
at the present time: e.g., cold gas accretion from filaments, gas dynamics, radiative cooling, star formation, stellar/AGN feedback, as well as all the possible 
effects induced by the interaction of galaxies with their surrounding environments. 

The distribution of galaxies in the Universe is highly inhomogeneous, with densities spanning several orders of magnitude. 
If $\rho_0$ is the average field density, the density varies from $\sim$ 0.2$\rho_0$ in voids
to $\sim$ 5$\rho_0$ in superclusters and filaments, $\sim$ 100$\rho_0$ in the core of rich clusters, up to 
$\sim$ 1000$\rho_0$ in compact groups (Geller \& Huchra 1989).
Although containing only $\sim$ 5\% ~ of the local galaxies, clusters are ideal laboratories to study the 
physical mechanisms perturbing galaxy evolution in dense environments.
Because of their high density, gravitational interactions between cluster members are expected to be frequent.
At the same time, clusters are characterised by a hot ($T$ $\sim$ 10$^7$-10$^8$ K) and dense ($\rho_{ICM}$ $\sim$ 10$^{-3}$ cm$^{-3}$)
intracluster medium trapped within their potential well (e.g. Sarazin 1986). The interaction of galaxies 
with this diffuse intracluster medium can easily remove their interstellar medium, thus affecting their star formation activity.

Environmental processes can be 
broadly separated in two classes: those related to the gravitational interactions between galaxies or with the potential 
well of over-dense regions (merging - Kauffmann et al. 1993; tidal interactions - Merritt 1983; Byrd \& Valtonen 1990; harassment - Moore et al. 1998), and 
those exerted by the hot and dense intracluster medium (ICM) on galaxies moving at high velocity within clusters (ram pressure 
stripping - Gunn \& Gott 1972; viscous stripping - Nulsen 1982; thermal evaporation - Cowie \& Songaila 1977; starvation - 
Larson et al. 1980). Since the large, dynamically-bounded structures observed in the local Universe form through the 
accretion of smaller groups of galaxies (Gnedin 2003; McGee et al. 2009; De Lucia et al. 2012), 
environmental processes are now believed to influence galaxies even before 
they enter rich clusters in the high-redshift Universe (pre-processing - Dressler 2004). This unexpected 
discovery has spawned a renewed interest in the detailed properties of galaxies in the local Universe, since such systems 
are the obvious test-beds for theories that attempt to explain this cosmic evolution. Indeed, studies of the 
high-redshift Universe can give us an integrated, statistical picture of galaxy evolution over cosmic time, but it is 
only through detailed studies of the local volume that we can hope to understand the detailed role of gas dynamics, 
cooling, star formation, feedback and environment in the hierarchical assembly of baryonic substructures.

Because of its proximity (16.5 Mpc), the Virgo cluster has been an ideal target for the study of the transformation of galaxies in rich environments.
The first blind photographic survey of the cluster, led by A. Sandage, G. Tammann and B. Binggeli, was possible only after 
the construction of the 2.5m (100-inch) Ir\'en\'ee du Pont telescope at Las Campanas (Chile) in 1977. 
The telescope was expressly designed to have an exceptionally wide field for direct photography (1.5$^o$ $\times$ 1.5$^o$) 
and was thus perfectly tuned to cover the whole Virgo cluster region. Only recently, the advent of the new generation of
large panoramic detectors made it possible to cover the whole cluster, which exceeds 100 deg$^2$, from the UV to the radio wavelengths. 

The studies of emission lines, however, which for wide field cameras require specific and expensive narrow-band filters
of large physical size, have thus far been limited to pointed observations.
Very deep H$\alpha$ observations of a few galaxies in nearby clusters, including our recent observations with MegaCam, 
have led to several intriguing discoveries. They have shown that the ionised phase appears to be an ideal 
tracer of stripped gas in dense regions: $\sim$ 50\% ~ of late-type galaxies show extended ($\sim$ 50 kpc) tails of ionised gas with surface 
brightness $\Sigma (H\alpha)$ $\sim$  a few 10$^{-18}$ erg sec$^{-1}$ cm$^{-2}$ arcsec$^{-2}$ (Boselli \& Gavazzi 2014), 
while only a handful of galaxies have extended cold or hot gaseous tails (Chung et al. 2007; Sun et al. 2006, 2007, 2010; Scott et al. 2012; 
Sivanandam et al. 2014; Jachym et al. 2014).
In some objects, the cometary shape of the tails indicates that the gas has 
been stripped by the interaction with the hot ICM (Gavazzi et al. 2001; Yoshida et al. 2002; Yagi et al. 2010; Fossati et al. 2012, 2016, 2018 - paper III; Zhang et al. 2013; Boselli et al. 2016a); 
in other systems, bridges of ionised gas linking different nearby galaxies 
are associated with tidal tails in the stellar component, suggesting gravitational perturbations with nearby companions or within infalling groups 
(i.e., pre-processing; Kenney et al. 2008; Sakai et al. 2012; Gavazzi et al. 2003a; Cortese et al. 2006).
They have also shown that within the tails of stripped gas, star formation in compact HII regions occurs in some but not in all objects (Gavazzi et al. 2001; Yoshida et al. 2008; Hester et al. 2010; 
Fumagalli et al. 2011b; Fossati et al. 2012; Boissier et al. 2012; Yagi et al. 2013; Kenney et al. 2014; Boselli et al. 2016a, 2018 - paper IV).
The removal of the gas affects the activity of star formation of galaxies on different timescales that depend on the 
perturbing mechanism (Larson et al. 1980; Boselli et al. 2006, 2016b; Bekki 2009, 2014; McGee et al. 2009; Cen 2014; Fillingham et al. 2015; Rafieferantsoa et al. 2015). 
The distribution and the morphology of the star-forming regions within galaxies is also tightly connected 
to the perturbing mechanisms (increases in the nuclear star formation activity and asymmetric distributions of star-forming 
regions are typical in gravitational interactions, radially truncated star-forming discs in interactions with the ICM, 
fainter star forming discs in starvation, Kennicutt \& Keel 1984; Barton et al. 2000; Boselli et al. 2006; 
Ellison et al. 2008; Scudder et al. 2012; Patton et al. 2011, 2013).
All these pieces of evidence underline the power of NB H$\alpha$ imaging data in identifying the dominant perturbing mechanism in dense environments.

\begin{table*}
\caption{The properties of the different Virgo cluster substructures.}
\label{structures}
{\tiny
\[
\begin{tabular}{ccccccccccccccc}
\hline
\noalign{\smallskip}
\hline
Substructure	& R.A.	& Dec.	 & Dist	& Ref	& $<vel>$	& $\sigma$	& Ref	 & $R_{200}$	& $M_{200}$   			& Ref	& $\rho_{ICM}$ 		& $T_{ICM}$ 	& Ref	& Central \\
	 & J2000 & J2000	&Mpc	& 	& km s$^{-1}$   &  km s$^{-1}$  &	 & Mpc		& M$_{\odot}$ $\times$10$^{14}$  & 	& cm$^{-3}$    		& keV	   	& 	& \\
\hline
Cluster A	& 187.71	& 12.39	 & 16.5	& 1,2 	& 955  	 	& 799  	 	& 3   	 & 1.55		& 1.4-4.2		   	& 4-9	& 2.0$\times$10$^{-3}$ 	& 2.3		& 10-11 & M87   \\
Cluster B	& 187.44	& 8.00	 & 23	& 1 	& 1134 	 	& 464  	 	& 3	 & 0.96		& 1			 	& 9	& 			& 		&	& M49   \\
Cluster C	& 190.85	& 11.45	 & 16.5	& 1 	& 1073 	 	& 545  	 	& 3	 & 0.66-1.15  	& 0.35-1.85			& 12-13	& 			&		& 	& M60   \\
W cloud		& 185.00	& 5.80	 & 32	& 1 	& 2176 	 	& 416  	 	& 3	 & 0.50-0.88	& 0.15-0.83			& 12-13	& 			&		&	& NGC4261	 \\
W' cloud	& 186.00	& 7.20	 & 23	& 1 	& 1019 	 	& 416  	 	& 3	 & 0.50-0.88  	& 0.15-0.83			& 12-13	& 			&		&	& NGC4365	 \\
M cloud		& 183.00	& 13.40	 & 32	& 1 	& 2109 	 	& 280  	 	& 3	 & 0.34-0.60	& 0.05-0.26			& 12-13	& 			&		&	& NGC4168	 \\
LVC cloud	& 184.00	& 13.40	 & 16.5	& 1 	& 85	 	& 208  	 	& 3	 & 0.25-0.44 	& 0.02-0.11			& 12-13	& 			&		&	& NGC4216	 \\
\noalign{\smallskip}
\hline
\end{tabular}
\]
References: 1) Gavazzi et al. (1999); 2) Mei et al. (2007); 3) Boselli et al. (2014); 4) Nulsen \& B\"ohringer (1995); 5) Girardi et al. (1998);
6) Schindler et al. (1999); 7) McLaughlin (1999); 8) Urban et al. (2011); 9) Ferrarese et al. (2012); 
10) B\"ohringer (2005), private communication; 11) B\"ohringer et al. (1994); 12) derived using the $M_{200}$ vs. $\sigma$ relation given in Biviano et al. (2006); 
13) derived using the $M_{200}$ vs. $\sigma$ relation given in Evrard et al. (2008).}
\end{table*}

   \begin{figure*}
   \centering
   \includegraphics[width=0.95\textwidth]{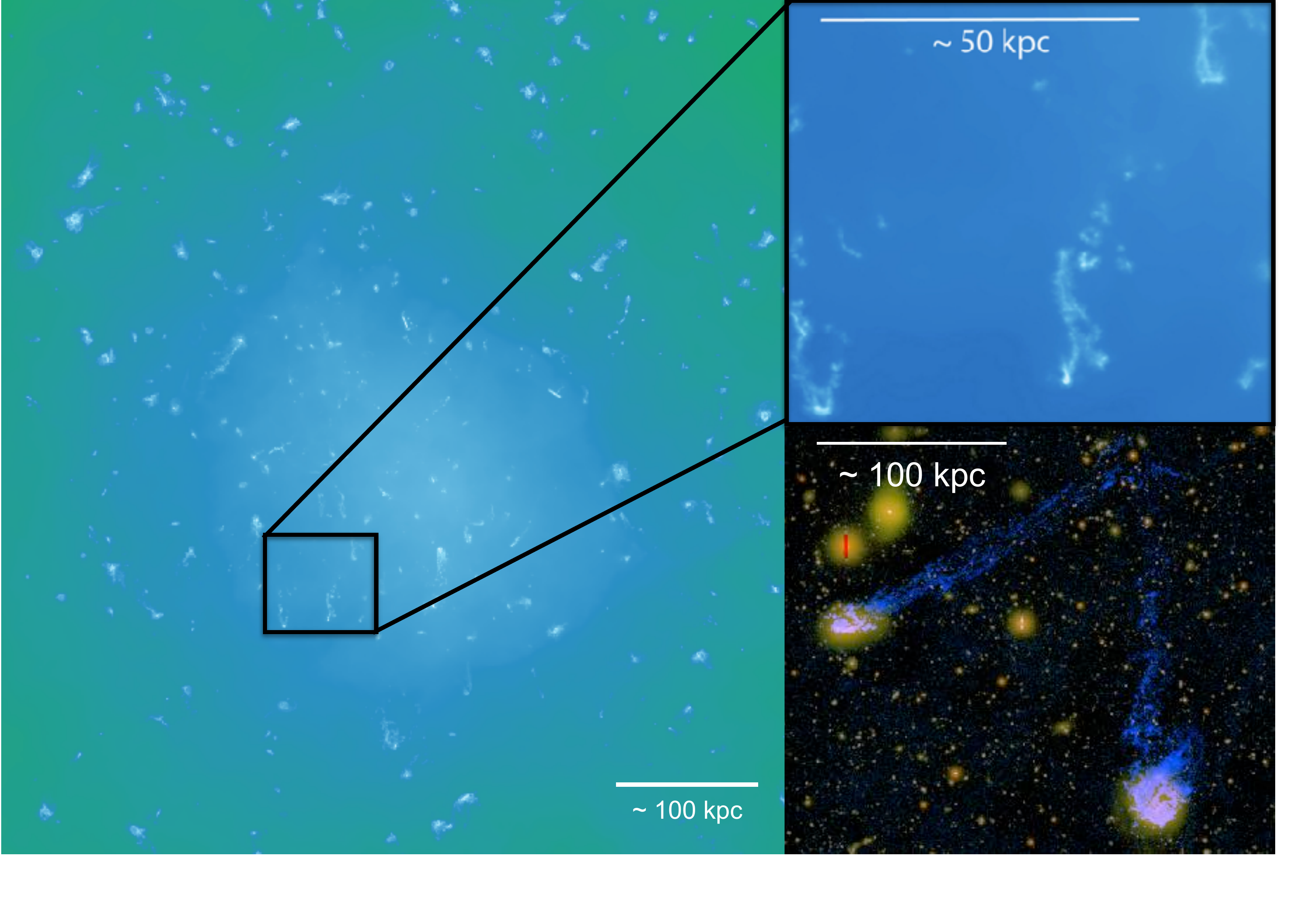} 
   \caption{Predictions for the gas distribution within a cluster of galaxies derived from the Illustris
   simulations (http://www.illustris-project.org/explorer; Nelson et al. 2015). Tails of stripped gas that are $\sim$ 50 kpc long, (upper right panel), 
   are expected to be associated with most of the simulated cluster galaxies. Once stripped from galaxies and injected into the hot 
   (10$^7$ - 10$^8$ K) ICM, the gas is ionised by heat conduction, turbulence, MHD waves and extraplanar star formation in the tails. 
   It then becomes visible in deep, wide-field NB H$\alpha$ imaging. Some gas associated to star forming regions 
   within the galactic disc can also be stripped in its ionised phase. Tails of ionised gas similar to those predicted by the simulations are indeed observed
   in the H$\alpha$ continuum-subtracted image of the two galaxies CGCG 97-73 and 97-79 in 
   A1367 obtained with Suprime-Cam at the Subaru telescope (10'$\times$10') (Boselli \& Gavazzi 2014; Yagi et al. 2017). Similar tails of ionised gas will be detected by
   VESTIGE.	  
   }
   \label{simulazioni}       
   \end{figure*}  
     
Many questions, however, remain unanswered due to the lack of systematic studies of the ionised diffuse gas through optical 
emission lines. This leaves many diagnostics that are key to the study of environmental processes 
nearly unexplored. Combined with multifrequency data, optical NB observations are crucial for identifying environment-induced 
effects on the different gaseous components and, ultimately, understanding the impact on star formation activity, stellar 
populations, and the fate of the stripped material. The main questions that a deep, complete, homogeneous H$\alpha$ survey of a 
nearby cluster will answer are:\\
\begin{enumerate}
\item What fraction of galaxies shows signs of perturbation as a function of galaxy mass and 
local density? What fraction of galaxies are perturbed by gravitational interactions or interactions with the ICM?\\
\item How efficiently is the gas stripped in the different phases during the interaction with the surrounding environment? How quickly 
does the gas change phase once it has been removed from the galaxy disc?\\
\item What are the ionising sources for the H$\alpha$ emitting 
gas (i.e., galactic or extraplanar HII regions, thermal conduction, turbulence, MHD waves, etc)?\\
\item How is the nuclear and disc 
star formation activity of galaxies perturbed during the interaction? At which point in the interaction is the star formation 
activity stopped: i.e., was the quenching rapid? \\    
\end{enumerate}

Within this framework, our team has been recently awarded 50 nights of telescope time at the Canadian French Hawaii Telescope (CFHT) 
to map the whole Virgo cluster region within one virial radius ($\sim$ 104 deg$^2$) with MegaCam using a newly commissioned
narrow-band H$\alpha$+[NII] filter\footnote{Hereafter we will refer to the H$\alpha$+[NII] band simply as H$\alpha$,
unless otherwise stated.}. This project, called VESTIGE (A Virgo Environmental Survey Tracing Ionised Gas 
Emission\footnote{http://mission.lam.fr/vestige/}), 
is one of the three large projects selected for the 2017-2019 observing campaigns at the CFHT. At the time of writing,
after the decommissioning of SUPRIME Cam from Subaru, there is no large telescope with wide-field capability 
other than the CFHT equipped with H$\alpha$ narrow-band imaging filters (at redshift 0 or above) anywhere in the world.
The principal aim of this survey is that of
studying the effects of the environment on galaxy evolution through the observation, the analysis, and the modelling
of the ionised gas phase of the interstellar medium (ISM) of galaxies stripped during their interaction with the hostile 
Virgo cluster. 

In this first paper we describe the survey design and the observing strategy, and we present 
the scientific motivations for this project. We also report on the status of the survey after the first semester of observations. 
The paper is structured as follows: in Sect. 2 we describe the 
Virgo cluster. In Sect. 3 we give the motivations, the instrumental set-up, the design of the survey and the observing strategy, in Sect. 4 the data 
processing and in Sect. 5 we explain how the full set of data will be made available to the community. In Sect. 6 we describe the observations
obtained in the 2017A semester for a 4$\times$1 deg$^2$ strip across the core of the cluster.
The scientific objectives of the survey are described in Sect. 7, while in Sect. 8 we present the synergy with other projects and we
list the planned follow-up observations necessary for a full exploitation of the data.
Throughout VESTIGE (unless otherwise stated in individual publications) we use a flat $\Lambda$CDM cosmology with $H_0$ = 70 km s$^{-1}$ Mpc$^{-1}$, $\Omega_m$ = 0.3, and
$\Omega_{\Lambda}$ = 0.7.

\section{The Virgo Cluster: A Unique Laboratory for Environmental Studies}    
 
The Virgo cluster is the richest cluster of galaxies within 35 Mpc. It is located at a distance of 16.5 Mpc 
(Gavazzi et al. 1999; Mei et al. 2007; Blakeslee et al. 2009) and has a total 
mass of $M_{200}$ = (1.4-4.2) $\times$ 10$^{14}$ M$_{\odot}$ (Nulsen \& B\"ohringer 1995, Girardi et al. 1998, 
Schindler et al. 1999, McLaughlin 1999, Urban et al. 2011; see Table \ref{structures}), where $M_{200}$ is the total mass within the radius in which 
the mean mass density is 200 times the critical cosmic mass density. 
With thousands of member galaxies lying at a nearly common distance and 
spanning all known morphological types, Virgo has historically played a key role in studies of how galaxies form and 
evolve in dense environments (e.g. Boselli \& Gavazzi 2006). Virgo is composed of a primary virialised system (subcluster A) dominated by 
quiescent early-type galaxies, and by several smaller substructures falling onto the main cluster and dominated by 
late-type systems (Binggeli et al. 1987; Gavazzi et al. 1999; Solanes et al. 2002; Boselli et al. 2014a). 
The physical properties of late-type galaxies vary dramatically from the 
periphery (where galaxies are virtually identical to unperturbed field objects in terms of gas content and star 
formation activity) to the cluster core (dominated by highly perturbed systems deprived of their gas and dust reservoir and 
with a significantly reduced star formation activity; Kennicutt 1983; Cayatte et al. 1990; Solanes et al. 2001; Vollmer et al. 2001; Gavazzi et al. 1998, 2002a, 2002b, 2005, 2006, 2013; 
Cortese et al. 2010a, 2012; Boselli et al. 2014a,2014c, 2016b). A wealth of observational evidence 
consistently indicates that Virgo is a young cluster still in formation (Tully \& Shaya 1984; Gavazzi et al. 1999; 
Karachentsev \& Nasonova 2010; Karachentsev et al. 2014; Sorce et al. 2016), and many individual member galaxies have 
been identified that highlight the various mechanisms by which environment can influence galaxy evolution
(e.g. Vollmer 2003; Kenney et al. 2004; Vollmer et al. 2004;   
Boselli et al. 2005, 2006, 2016a; Haynes et al. 2007; Crowl \& Kenney 2008; Abramson et al. 2011). Virgo 
is thus an ideal laboratory 
for studying (at high resolution) the perturbing mechanisms that shaped galaxy evolution.

The development of 
wide-field, ground- and space-based facilities has made Virgo accessible for blind surveys at different wavelengths, 
from the X-ray to the UV, visible, near- and far-IR, and radio, allowing astronomers to map at exquisite 
sensitivity and angular resolution the different constituents of galaxies (e.g., stars, cold gas in the atomic and molecular 
phase, ionised and hot gas, dust, magnetic fields) and the intracluster medium (see Sect. 8). At the distance of Virgo, the typical 1-10" 
resolution achieved by these surveys corresponds to $\sim$ 0.1-1 kpc and thus perfectly matches that of the most recent cosmological
hydrodynamic simulations. Moreover, the dwarf galaxy population is accessible down to $M_{star}$ $\sim$ 10$^5$ M$_{\odot}$ 
(NGVS, Ferrarese et al. 2016; Roediger et al. 2017). Virgo is thus the ideal target to extend the stellar mass dynamic range sampled by SDSS and other 
local surveys by nearly two orders of magnitude. Since it is spiral-rich, it is also better suited than more distant relaxed 
clusters, such as Coma, for identifying galaxies being transformed by their environments. Furthermore, Virgo has physical properties 
(dynamical mass, gas temperature and density) significantly different than those encountered in other nearby clusters
such as Coma, A1367, and Norma, and thus can be used to extend previous studies to less extreme but more frequent 
and representative over-dense regions in the local Universe. For these reasons Virgo is, 
without question, the most thoroughly studied cluster of galaxies in the Universe, and remains the best target at low-redshift 
for a systematic study of the different perturbing mechanisms acting on galaxies in dense environments.

   \begin{figure}
   \centering
   \includegraphics[width=9.5cm]{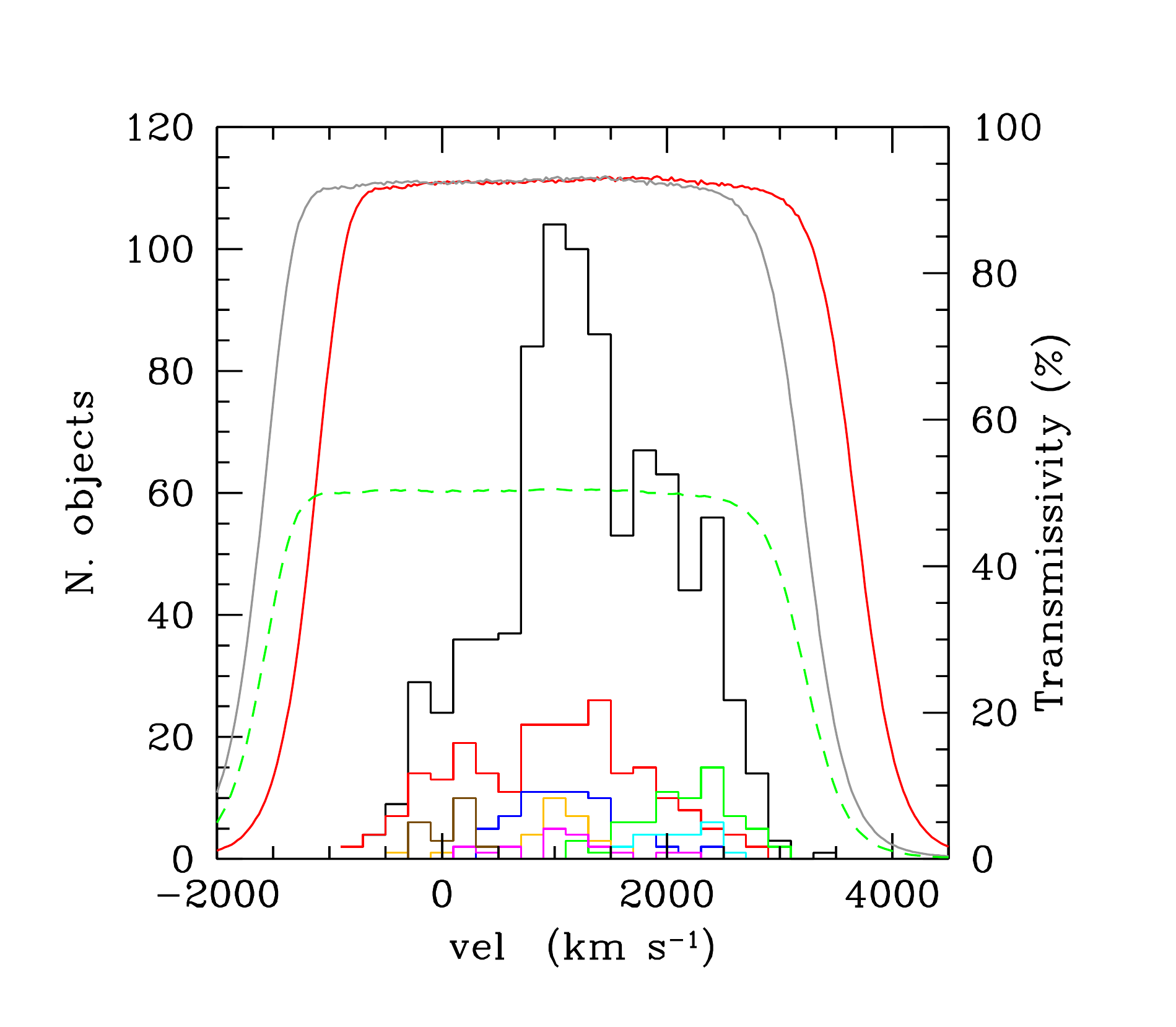}
   \caption{The velocity distribution of galaxies within the VESTIGE footprint (black histogram) 
   is compared to the transmissivity of the H$\alpha$ NB filter as measured in laboratory (red solid line)
   or expected for the typical spring observing conditions ($T$ = 0$^o$C; grey solid line). The green dashed line 
   shows the combined trasmission for mirrors, optics, filter, and detectors. The velocity distribution 
   of galaxies located within the different subclouds of the cluster defined as in Boselli et al. (2014a) 
   are given with the coloured histograms: red - cluster A, blue - cluster B, green - W cloud, orange - W' cloud,
   cyan - M cloud, magenta - cluster C, brown - low velocity cloud. 
 }
   \label{vel}%
   \end{figure}

\section{VESTIGE: A Deep H$\alpha$ Survey of the Virgo Cluster}    
 
A systematic H$\alpha$ survey of Virgo is imperative if we are to understand 
the effects of the environment on galaxy evolution. VESTIGE has been designed to detect the 
low surface brightness tails of ionised gas observed in a handful of nearby cluster galaxies or predicted by hydrodynamic 
simulations to be the smoking gun of ongoing gas stripping (Tonnesen \& Bryan 2010). 
VESTIGE will thus produce the definitive dataset against which to compare and test cosmological models 
of galaxy formation (Fig. \ref{simulazioni}). Compared to previous observations, which have so far targeted only a few, subjectively-selected 
galaxies, VESTIGE will provide dramatic improvements in depth ($\sim$100$\times$ in luminosity for point sources), surface brightness sensitivity ($\sim$30$\times$), 
angular resolution ($\sim$3$\times$ in seeing), and sky coverage ($\sim$100$\times$). Equally important, the survey will open 
many synergistic opportunities with planned and ongoing Virgo surveys at other wavelengths.

\subsection{CFHT and MegaCam}

The observations are carried out using MegaCam, a wide-field optical imager mounted on the prime focus 
of the 3.6 m CFHT (Boulade et al. 2003). The focal ratio of the camera is F/3.77. MegaCam is composed of 40 back-illuminated 2048$\times$4096 pixels CCDs,
with a pixel size of 0.187 arcsec on the sky and a typical read noise of $\sim$ 5 $e$ pixel$^{-1}$. Their typical efficiency at 6563 \AA ~ is 77\%. 
The inner 36 CCDs cover a rectangular field of view of size 0.96 $\times$ 0.94 deg$^2$, while the remaining (unvignetted) CCDs are located at the R.A. 
edges of the camera (see Fig. 3 in Ferrarese et al. 2012). The gaps between the different CCDs are either 13" or 80" wide due to the camera design.

\subsection{The narrow-band H$\alpha$ filter}

The observations employ the newly commissioned H$\alpha$ NB filter (MP9603). The filter bandpass measured for a F/8 focal ratio at a
temperature of 20$^o$C covers the range 6538 $<$ $\lambda$ $<$ 6644 \AA ~(central wavelength $\lambda$ = 6591\AA, 
$\Delta\lambda$ 106 \AA, with a typical transmissivity of 93\%). 
The transmissivity curves of NB interferential filters slightly 
changes with focal ratio and temperature. A blueshift of the central wavelength is expected when going from a parallel to 
a converging beam (the shift is non linear), with a typical blueshift of 0.25 \AA/degree is expected for decreasing temperature. A decrease of the focal ratio and of the
temperature also induce a very small decrease of the peak transmissivity and a slight broadening of the filter. Given that the observations are carried out 
in spring time, the typical temperature at the telescope is $T$ $\sim$ 0$^o$C, corresponding to $\sim$ 5 \AA ~ blueshift. Adding 
an extra $\sim$ 5 \AA ~ blueshift for changing the focal F/8 to F/3.77, the expected blueshift of the central wavelength during observing conditions 
is of $\sim$ 10 \AA. If we do not consider any extra marginal variation of the transmissivity curve, the NB filter covers the velocity range 
-1140 $<$ $cz$ $<$ +3250 km s$^{-1}$ for the H$\alpha$ line, and is thus perfectly suited for the velocity range of galaxies within the cluster (Binggeli et al. 1987;
Boselli et al. 2014a), as depicted in Fig. \ref{vel}. The filter width also brackets the two [NII] lines at $\lambda$ 6548 \AA ~ and 6583 \AA.

\subsection{Survey geometry, exposure times and achieved depth}

\begin{figure*}
   \centering
   \includegraphics[width=16cm]{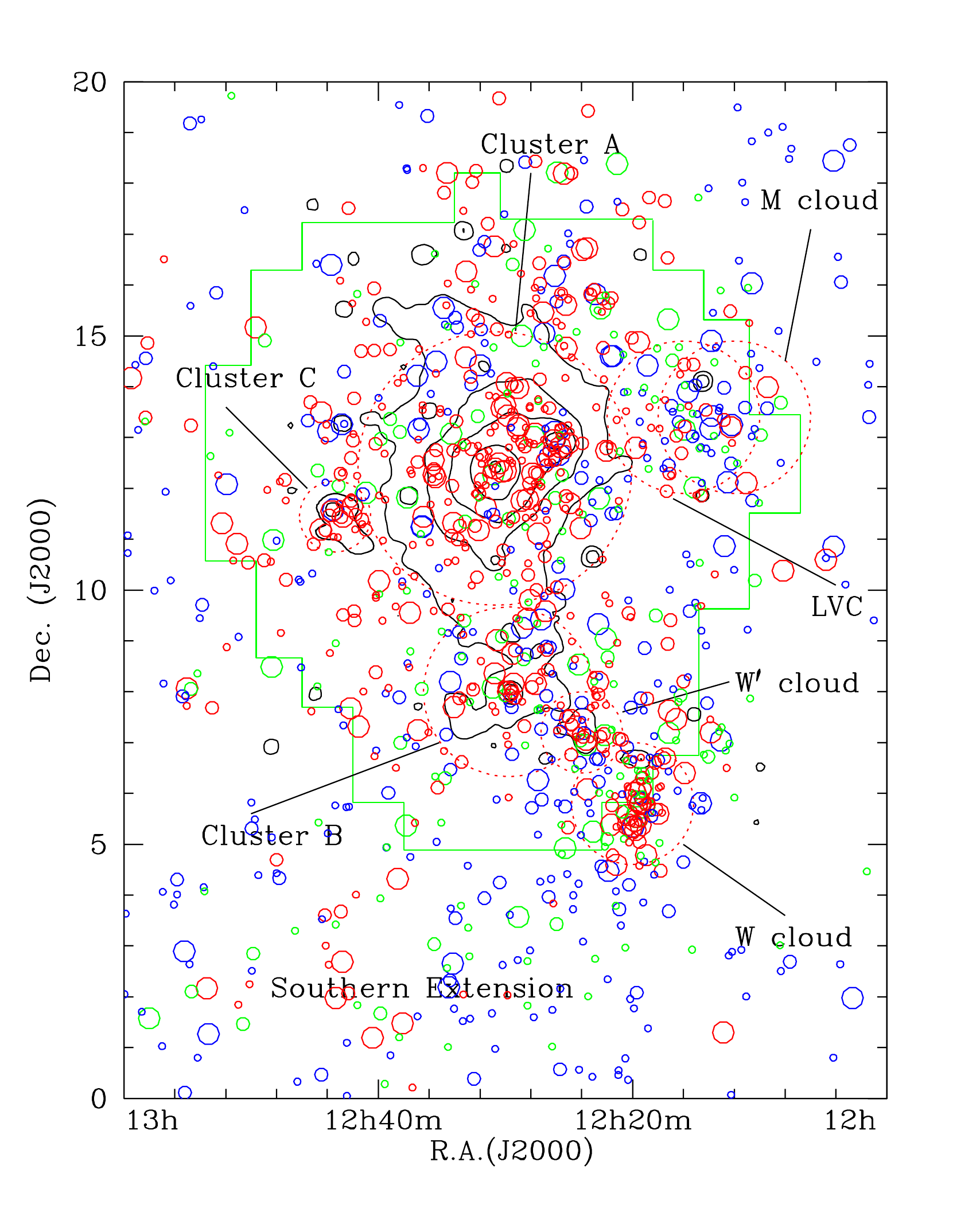}
   \caption{The Virgo cluster region mapped by VESTIGE. The complete blind survey covers the inner 104 deg$^2$
   (green footprint), the same region mapped by NGVS (Ferrarese et al. 2012). The black contours indicate the X-ray emission of 
   the diffuse gas of the cluster obtained by ROSAT (B\"ohringer et al. 1994), the large red dotted circles the different substructures 
   of the cluster, while the red, green, and blue empty circles early-type galaxies, transition type galaxies, and star forming systems 
   respectively, with sizes depending on their stellar mass (big for $M_{star}$ $>$ 10$^{9.5}$ M$_{\odot}$, 
   medium for 10$^{8.5}$) $<$ $M_{star}$ $\leq$ 10$^{9.5}$ M$_{\odot}$, small for $M_{star}$ $\leq$ 10$^{8.5}$ M$_{\odot}$) as 
   defined in Boselli et al. (2014a). }
   \label{angdist}
   \end{figure*}

Virgo is located between 12h$<$ R.A. $<$ 13h and 0$^o$ $<$ dec $<$ 18$^o$. 
NGVS catalogued $\simeq$ 3700 Virgo cluster member galaxies within the VESTIGE footprint, 
out of which $\sim$ 550 are blue, presumably star-forming systems (Ferrarese et al., in prep.)
as depicted in Fig. \ref{angdist}. 


The primary goal of VESTIGE 
is the detection of low surface brightness features such as those observed in a few cluster galaxies (Gavazzi et al. 2001; Sun et al. 2007; 
Kenney et al. 2008; Yagi et al. 2010; Fossati et al. 2012; Boselli et al. 2016a) or predicted by hydrodynamic simulations (Roediger \& Hensler 2005; Tonnesen \& Bryan 2009, 2010). 
Such features have a typical angular size of $\sim$ 20 arcmin and a surface brightness of 
$\Sigma (H\alpha)$ $\sim$ 1-2 $\times$ 10$^{-18}$ erg s$^{-1}$ cm$^{-2}$ arcsec$^{-2}$ and thus require a sensitivity 
$\sim$ 60 $\times$ fainter than those reached in previous observations of Virgo galaxies using 2m class telescopes. 

The total integration time of VESTIGE was determined after the pilot obeservations of NGC 4569 done in 2015 (Boselli et al. 2016a).
The total integration time was set to 7200 sec and of 720 sec in the NB and broad-band $r$ filters, respectively. 
Observations in the broad-band filter are necessary for the subtraction of the stellar continuum. The integration time 
in the broad-band, which is $\sim$ 14$\times$ wider than the NB, has been chosen to reach approximately the same sensitivity.
The observations carried out at the CFHT during the 2017A semester in dark/grey time have shown that the typical sensitivity of VESTIGE is 
$\Sigma (H\alpha)$ $\sim$ 1.5 $\times$ 10$^{-17}$ erg s$^{-1}$ cm$^{-2}$ arcsec$^{-2}$ (1$\sigma$) at full resolution (0.187 arcsec pixel), 
and $\Sigma (H\alpha)$ $\sim$ 2 $\times$ 10$^{-18}$ erg s$^{-1}$ cm$^{-2}$ arcsec$^{-2}$ (1$\sigma$) 
once the data are smoothed to an angular resolution of $\sim$ 3 arcsec suitable for the detection of extended sources. 
The images gathered in semester 2017A have been also used to estimate empirically  
the 90\% completeness limit (5$\sigma$ level) for point
sources. Our limiting magnitudes were computed via simulations of point-like populations,
modeled with a Gaussian profile with FWHM=0.7\arcsec, in the
magnitude range $23.0 \le m_{\mathrm{AB}} \le 28.0$, and randomly
distributed on the images. We then carried out the
photometry as for the real sources with SExtractor and analysed, as
function of the magnitude, the recovery fraction of the synthetic
population for the subsample of objects characterised by S/N$\ge
5$. The limit of completeness was then defined as the magnitude for
which 90\% of the input objects were retrieved. We found
$m(NBH\alpha)_{\mathrm{lim}}= 24.4$ AB mag, corresponding to
$f(H\alpha)_{\mathrm{lim}}$ = $4\times 10^{-17}$ erg s$^{-1}$ cm$^{-2}$, and
$m(r)_{\mathrm{lim}}= 24.5$ AB mag, for the narrow- and broad-band
images, respectively (see Table~\ref{sensitivity}).
The sensitivity for point sources is close to the detection limit for PNe, 
that in Virgo is at $f(H\alpha)$ $\sim$ 3 $\times$ 10$^{-17}$ erg s$^{-1}$ cm$^{-2}$ as derived from their H$\alpha$ luminosity 
function (Ciardullo 2010), and is sufficient to sample the bright end of the $z$=4.4 Ly$\alpha$ luminosity function
which has a characteristic flux of $f(Ly\alpha)$ = 2-4 $\times$ 10$^{-17}$ erg s$^{-1}$ cm$^{-2}$ (Ouchi et al. 2008; Cassata et al. 2011). 


To conclude, each sky position is observed with a 7200s integration in H$\alpha$  
and 720s integration in the $r$-band. 
Each exposure, in both H$\alpha$ and $r$, is divided in 12 exposures optimally dithered to 
cover the CCD gaps and minimise the possible contamination from the reflections of bright stars (see below).

\begin{table*}
\caption{Sensitivity of VESTIGE.}
\label{sensitivity}
{
\[
\begin{tabular}{ccccc}
\hline
\noalign{\smallskip}
\hline
Band			& point source		& units					&	extended sources	& units	\\
			& 			&					&				&	\\
\hline
 $r$ (MP9602)    	& 	24.5		& AB mag (5$\sigma$)			& 25.8				& AB mag arcsec$^{-2}$ (1$\sigma$)\\
 NBH$\alpha$ (MP9603)	& 	24.4		& AB mag (5$\sigma$)			& 25.6				& AB mag arcsec$^{-2}$ (1$\sigma$)\\
 $f(H\alpha)$   	& 4$\times$10$^{-17}$ 	& erg s$^{-1}$ cm$^{-2}$ (5$\sigma$)	& 2$\times$10$^{-18}$ 	& erg s$^{-1}$ cm$^{-2}$ arcsec$^{-2}$ (1$\sigma$)$^a$\\
\noalign{\smallskip}
\hline
\end{tabular}
\]
Note: $a$ after smoothing the data to $\sim$ 3 arcsec resolution.}
\end{table*}

\subsection{Data acquisition and observing strategy}


   \begin{figure*}
   \centering
   \includegraphics[width=0.95\textwidth] {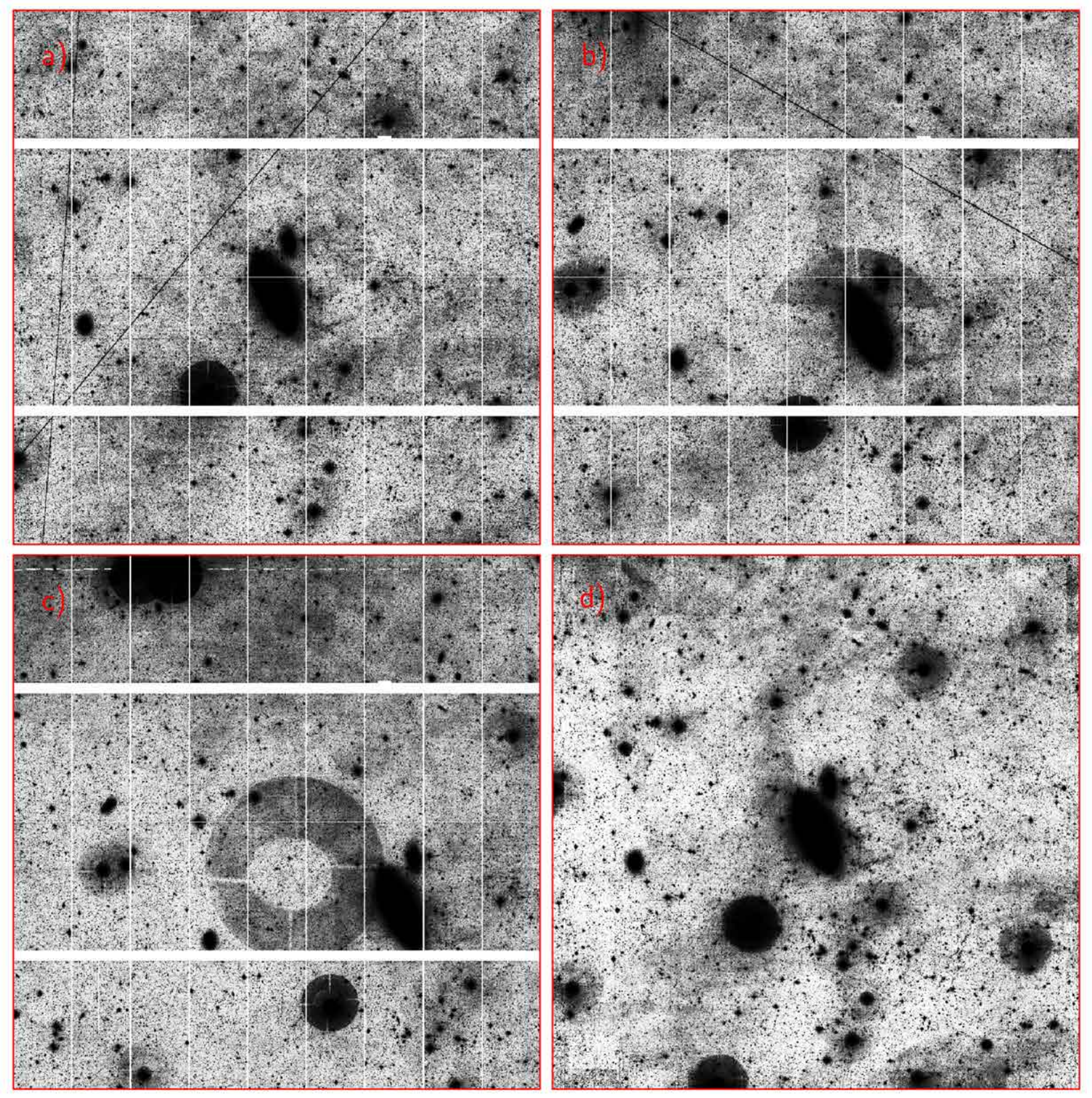}
   \caption{Panels a, b, and c: MegaCam images of the galaxy NGC 4569 obtained during three different
   pointings reduced using the standard Elixir pipeline. A prominent ghost due to the reflection of a star is evident in two of them 
   (panels b and c). Given their different position on
   the frame, the ghosts are efficiently removed after stacking seven independent frames reduced using Elixir-LSB (panel d; the grid pattern in the periphery 
   of this frame is due to undersampled regions). 
 }
   \label{reflections}%
   \end{figure*}

\begin{figure*}
   \centering
   \includegraphics[width=1.0\textwidth] {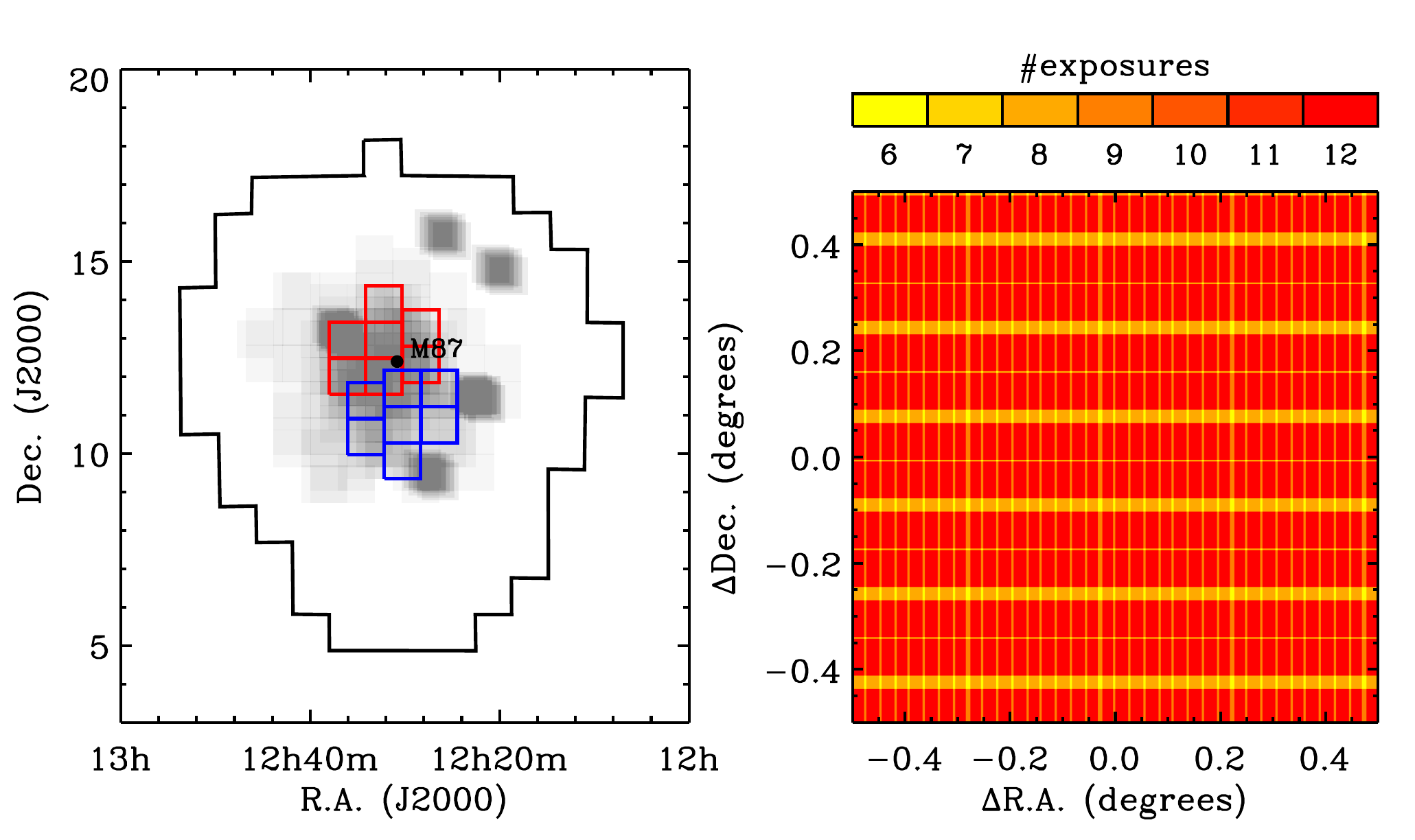} \\
   \caption{Left: VESTIGE maps of the 104 deg$^2$ within the NGVS footprint with 1419 independent frames in both the 
   H$\alpha$ NB and the $r$-band filters. To optimise the flat-fielding using the Elixir-LSB pipeline, 
   the observations are gathered within an observing block of seven concatenated and adjacent frames, as depicted by the red and blue footprints here taken as twe examples.
   The total number of observing blocks requested to cover the whole cluster is 203. Those obtained after the 2017A observing campaign are marked in 
   grey with brightness increasing with the number of completed exposure (0 = white, 12 = dark grey). The dark regions in the outskirts of the clusters 
   are the pilot observations completed in 2015 and 2016.
   Right: each
   sky region is covered by 12 independent frames dithered by 20 arcmin in R.A. and 15 arcmin in Dec. This large dithering secures 
   the sampling of the gaps between the different CCDs of MegaCam with a minimum of six exposures.
 }
   \label{pattern}
   \end{figure*}

The data are acquired following 
the Elixir-LSB pointing strategy successfully developed for the NGVS and optimised for the best characterisation of the sky background and 
the detection of extended low surface brightness features. The strategy requires an uninterrupted sequence of seven single exposures 
of contiguous fields to minimise any variation in the background illumination. The observations taken during a pilot project have shown that 
in the NB flat fielding can be severely affected by the reflection of bright stars ($\sim$ 7 mag) 
both within and outside the MegaCam field of view (see Fig. \ref{reflections}). These reflections are more important in the NB than in the broad-band filters.
The bright stars located within the MegaCam field produce circular annuli around the star with properties depending on the brightness and on
the position of the star within the field. Since these reflections depend on the optical configuration of the camera, the MegaCam
community is trying to model these reflections with the purpose of defining and producing dedicated pipelines able to remove them from the images
(Regnault et al. in prep.). The reflections due to stars outside the MegaCam field of view produce ghosts 
that change significantly according to the relative position of the camera and of the star. These ghosts can be easily removed using a median 
stacking of dithered images provided that each single frame is taken at significantly different positions (see Fig. \ref{reflections}). 
We thus adopt an observing pattern defined to cover the same sky region in 12 different frames taken after a large dithering of 15 arcmin in R.A. 
and 20 arcmin in Dec. Each single exposure is of 600s in the NB and 60s in the $r$-band. To map the full NGVS footprint VESTIGE requires
1419 single pointings in each filter, or equivalently 203 observing blocks of 7 frames each. This particular observing pattern 
covers the gaps between the different CCDs composing MegaCam. With this pattern, most of the Virgo cluster will be 
covered by 12 independent frames, while only a small fraction coinciding with the gaps of the CCDs by 
6, 8, 9 or 10 frames, as depicted in Fig. \ref{pattern}. 


Pilot observations undertaken in 2015 and 2016 indicate that with single exposures of 600s in the NB and 60s in the $r$-band the frames are background limited in both
bands ($\simeq$ 195 ADUs in $r$ and 141 ADUs in NB taken in dark time). 
Short exposures (60s in 
H$\alpha$ and 6s in $r$) might be necessary whenever the analysis of the deep images reveals saturation in the nucleus of bright galaxies. Based on 
the NGVS data, we expect $\sim$ 80 galaxies to saturate in the long exposures. 

During the spring 2017 observing campaign (semester 2017A) 244 pointings in the $r$-band (17\%~ of the full survey) and 268 (19\%) in the NB filter have been acquired. 

\subsection{Image quality} 

\begin{figure}
   \centering
   \includegraphics[width=9.5cm] {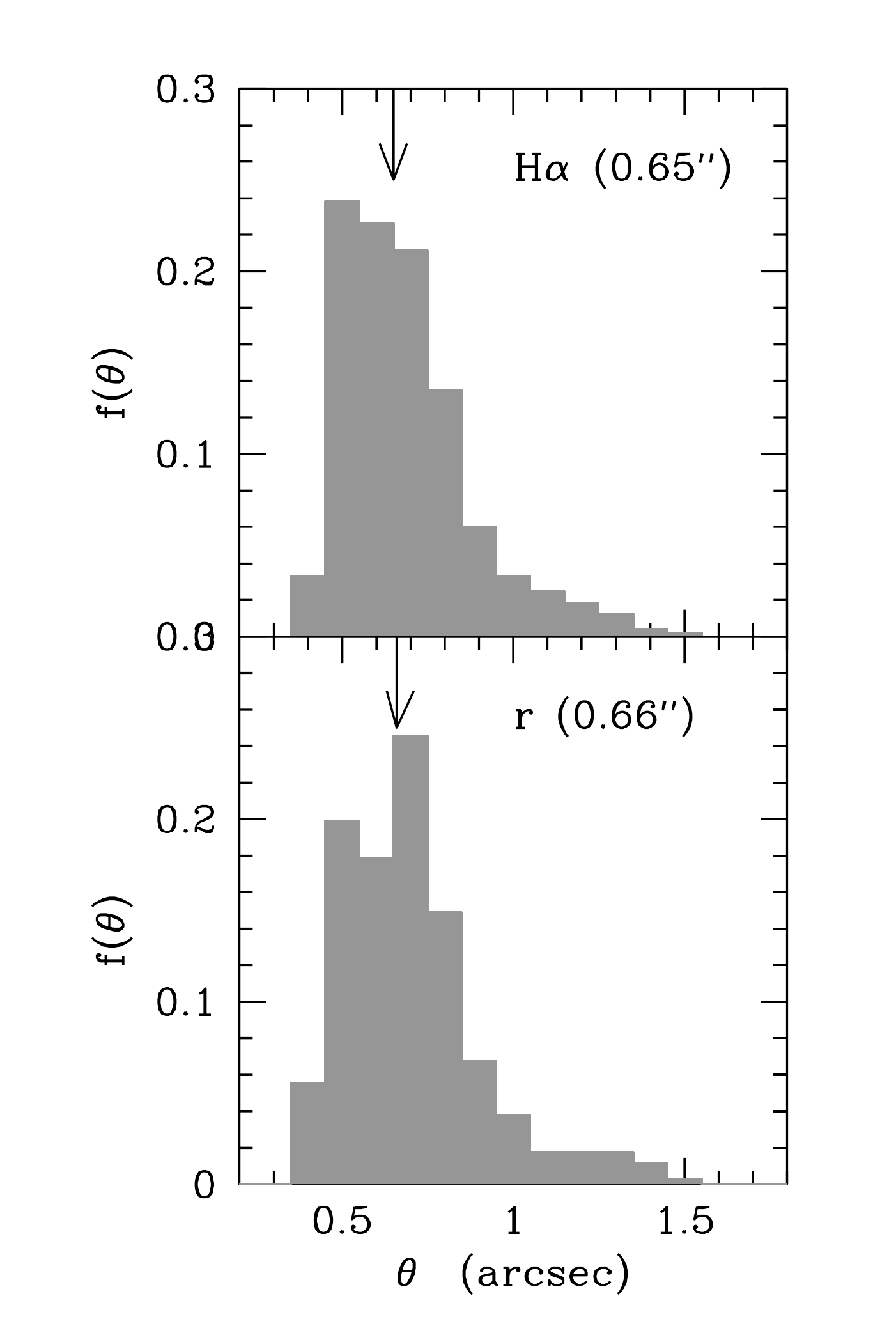} \\
   \caption{Seeing distribution in the H$\alpha$ NB (upper panel) and $r$-band (lower panel) filters determined
   from the single images gathered during the 2017A observing campaign and in the 2015 and 2016 pilot observations. The vertical arrows indicate the median 
   values.
 }
   \label{seeing}%
   \end{figure}

Given the extended, low surface brightness nature of the ionised gas tails, no strong constraints on the seeing conditions are needed for the survey. 
However, given the exceptional imaging quality of the CFHT, the observations will be gathered in subarcsecond conditions. Figure \ref{seeing}
shows the seeing distribution for the single exposures in the H$\alpha$ and $r$-band filters obtained during the 2017A semester and 2015 and 2016 pilot observations.
The median seeing in the two filters is 0.64 and 0.65 arcsec, respectively.


\section{Data processing}

VESTIGE benefits from the extensive machinery developed to support the NGVS survey completed with the same instrument. 
Accurate flat-fielding is done using Elixir-LSB, while
global astrometric solutions, image stacking and source catalogues are generated by a modified version of MegaPipe (Gwyn 2008).

\subsection{Elixir-LSB}
 
The MegaCam images are reduced with Elixir-LSB (Ferrarese et al. 2012), an upgrade of the Elixir (Magnier \& Cuillandre 2004)
pipeline specifically developed within the NGVS collaboration to detect extended low surface brightness features associated to cluster galaxies,
whose scales of $\lesssim$ 20 arcmin do not exceed the MegaCam field of view. This specific pipeline has been designed to 
remove any possible residual large scale structure in the sky background, including the low-level diffuse Galactic emission extending all over the VESTIGE footprint
(see sect. 7.2.1).
Elixir-LSB can be applied whenever the different frames are acquired 
under a similar background illumination, i.e. within a concatenated cycle of seven independent exposures (see Sect. 3.4). The images must be background dominated,
as is the case for the VESTIGE data in both the broad- and NB filters.  
The Elixir-LSB pipeline performs bad pixel masking, bias and overscan corrections, and flat-fielding.

\subsection{Astrometric calibration}

While the MegaPipe pipeline now uses GAIA (Perryman et al. 2001) as an astrometric reference
frame, for consistency with NGVS the VESTIGE data were astrometrically calibrated
using the NGVS data (Ferrarese et al. 2012) as a reference, the latter being calibrated on the SDSS. There are
small but significant astrometric shifts relative to GAIA, on the
order of 100 mas or less. The internal astrometric residuals, on the
other hand, are typically 20-40 mas.

\subsection{Photometric calibration}

The VESTIGE data is bootstrapped from Pan-STARRS PS1 photometry
(Magnier et al. 2013) which is accurate to within $\leq$ 0.005 mag
(Finkbeiner et al. 2016) and is sufficiently deep to provide a large number of
common stars per field to minimise random errors per source. We first apply a $\sim$ 0.2 mag 
flat field radial correction from the centre to the edges to take into account the different transmissivity 
of the camera. This correction has been determined after the repeated observation of a nearby cluster of stars
observed in different positions across the field.
The magnitudes of stars in each VESTIGE
image are transformed from the PS1 filter system into the MegaCam
filter system and then used as in-field standards. The transformations
are derived by first computing the transmission functions of the PS1
and MegaCam passbands (including the reflectance of the primary
mirrors, the camera optics, the quantum efficiency of the CCDs and the
transmittance of the filters themselves). These total transmission
functions are multiplied by stellar spectra from both the HST
CALSPEC spectra collection\footnote{http://www.stsci.edu/hst/observatory/crds/calspec.html} 
and Pickles (1998).

The transformations are as follows:

\begin{eqnarray}
r_{\rm MP9602} = r_{\rm PS1} +0.00020-0.01700 x\\
+0.00554 x^2 -0.000692 x^3 \nonumber \\
H\alpha_{\rm MP9603} = r_{\rm PS1} +0.08677-0.29983 x\\
+0.15859 x^2 -0.055190 x^3 \nonumber
\label{eqn:trans}
\end{eqnarray}

\noindent
where $x=g_{\rm PS1}-r_{\rm PS1}$.

The MegaCam instrumental magnitudes are measured through circular
apertures that increase in size with seeing. The aperture is such
that, for point sources, the measured flux is equivalent to that
measured by a Kron magnitude such as SExtractor's
$\tt{MAG\underline{~}AUTO}$.

For each exposure, a single zero-point is determined for the entire
mosaic. In practice there are small variations in the zero-point
across the focal plane ($\lesssim$ 0.03 mag), however because of the nature of the low surface brightness
background subtraction, these are not corrected. The zero-points are
determined for each exposure independently. Comparing photometry from
overlapping images indicates that the zero-points are self-consistent
to about 0.01 magnitudes. We also derived a set of relations similar to eq.
(1-2) using a procedure to transform SDSS magnitudes into MegaCam
magnitudes to check the consistency on the photometry between the two surveys. 
The SDSS zero-points were consistent with the PS1-derived zero-points of VESTIGE
to within 0.01 magnitudes.

\subsection{Stacking}

The calibrated images are resampled on to pixel grid matching the NGVS
image footprints using
SWarp\footnote{http://www.astromatic.net/software/swarp} and the
astrometric solution. The images are scaled according to the photometric
solution. The resampled, scaled images are combined using an artificial
skepticism algorithm (Stetson
1987)\footnote{http://ned.ipac.caltech.edu/level5/Stetson/Stetson\underline{~}contents.html}.
As with  the NGVS data
products, the resulting stacks have a zero point of 30.0 mag, such that AB
magnitudes are given by:

\begin{equation}
m_{\rm AB} = -2.5 \times \log(CNTS) + 30.0
\end{equation}

\subsection{H$\alpha$ fluxes and equivalent widths}

H$\alpha$ fluxes and equivalent widths are determined following standard procedures such as those described in Kennicutt et al. (2008). 
Given the depth of the present survey, the accuracy on the H$\alpha$ flux determination depends on an accurate subtraction of the stellar continuum.
Given the width of the $r$ filter
($\lambda_c$ = 6404 \AA; $\Delta \lambda$ = 1480 \AA) and the slight difference in the peak wavelength of the two bands, 
the derivation of the stellar continuum in the narrow-band from the $r$-band depends on the spectral properties of the 
emitting source (Spector et al. 2012). Using several hundred thousands of unsaturated stars detected in the science frames observed
during the pilot observations and in the 2017A observing run we were able to calibrate an empirical relation 
between the colour of the stars and the normalisation factor:

\begin{equation}
{\frac{r}{H\alpha} = r - 0.1713 \times (g-r) + 0.0717}
\end{equation}

We then apply this normalisation pixel by pixel on the stacked frame before the subtractioon of the stellar continuum. 
The $g-r$ colour map of any target can be derived using the $g$-band frame taken with MegaCam during the NGVS survey (Ferrarese et al.
2012). To avoid the introduction of any extra noise in the sky regions, where there is no stellar continuum, this colour-dependent normalisation is applied
only whenever the signal in the $r$- and $g$-bands has a signal-to-noise above a given threshold. 
The methodology used to derive fluxes and equivalent widths from the VESTIGE NB images will be extensively described in a dedicated publication 
(Fossati et al., in prep., paper II).



\section{Data access}

All data and data products will be stored at, and distributed by, the Centre of Astrophysical Data in Marseille (CeSAM)
through a dedicated web-page (http://mission.lam.fr/vestige/), 
as successfully done for several 
projects to which some team members are leaders (The \textit{Herschel} Reference Survey http://hedam.lam.fr/HRS/; 
GUViCS http://galex.lam.fr/guvics/index.html; NGVS http://www.cadc.hia.nrc.gc.ca/en/community/ngvs and GOLDMine http://goldmine.mib.infn.it/).
The stored data will include the fully reduced, stacked images in both the NB and $r$-band, as well as the continuum-subtracted frames.
It will also include the full catalogues of point sources in both bands. 
 

\section{First results}

\subsection{Comparison with previous observations}

\begin{figure*}
   \centering
   \includegraphics[width=13cm] {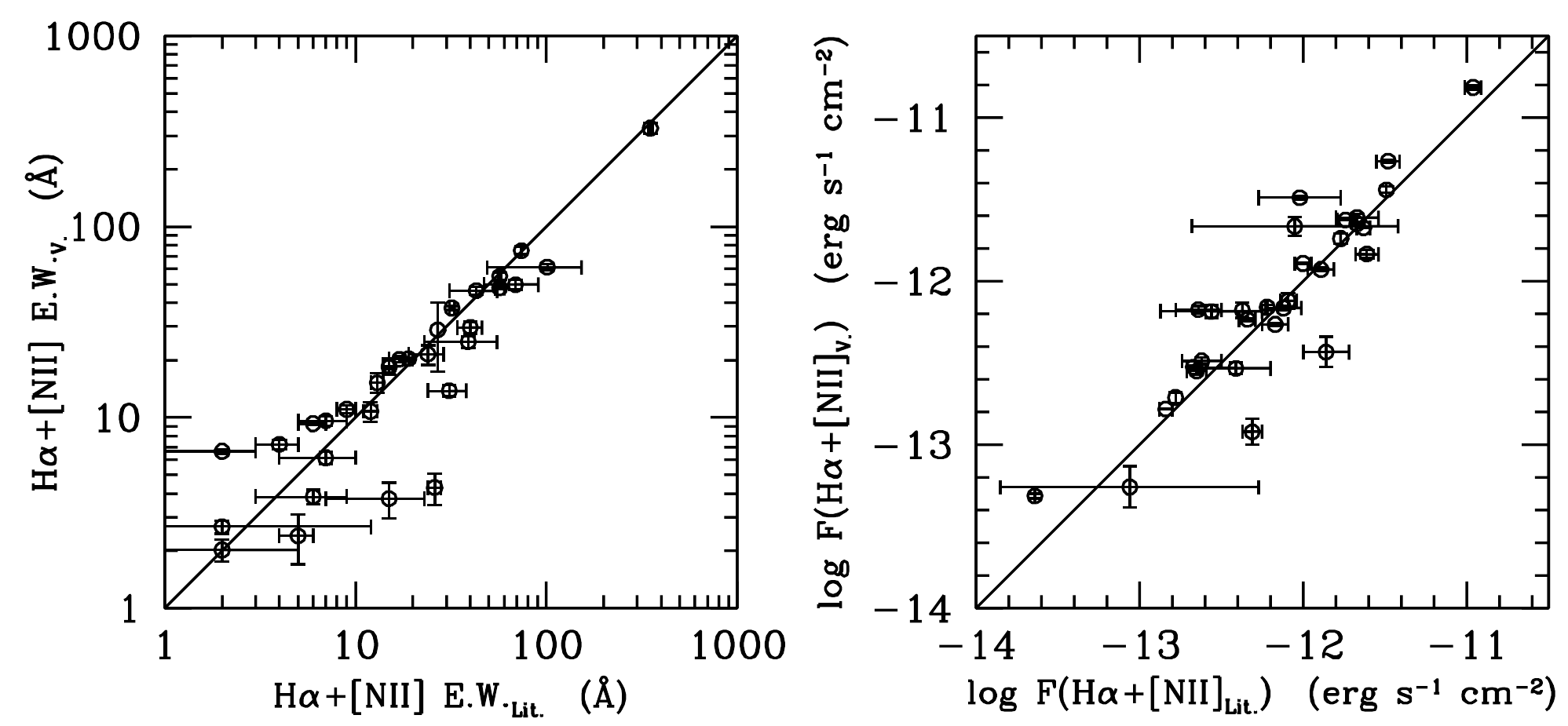} \\
   \caption{Comparison of the H$\alpha$ equivalent width (left) and fluxes (right) of the Virgo galaxies observed during the first 2017A semester of the VESTIGE survey 
   or during the pilot projects (Y-axis) with independent measurements available in the literature (X-axis). The solid line shows the 1:1 relation.
 }
   \label{comp}%
   \end{figure*}

\begin{figure}
   \centering
   \includegraphics[width=9cm]{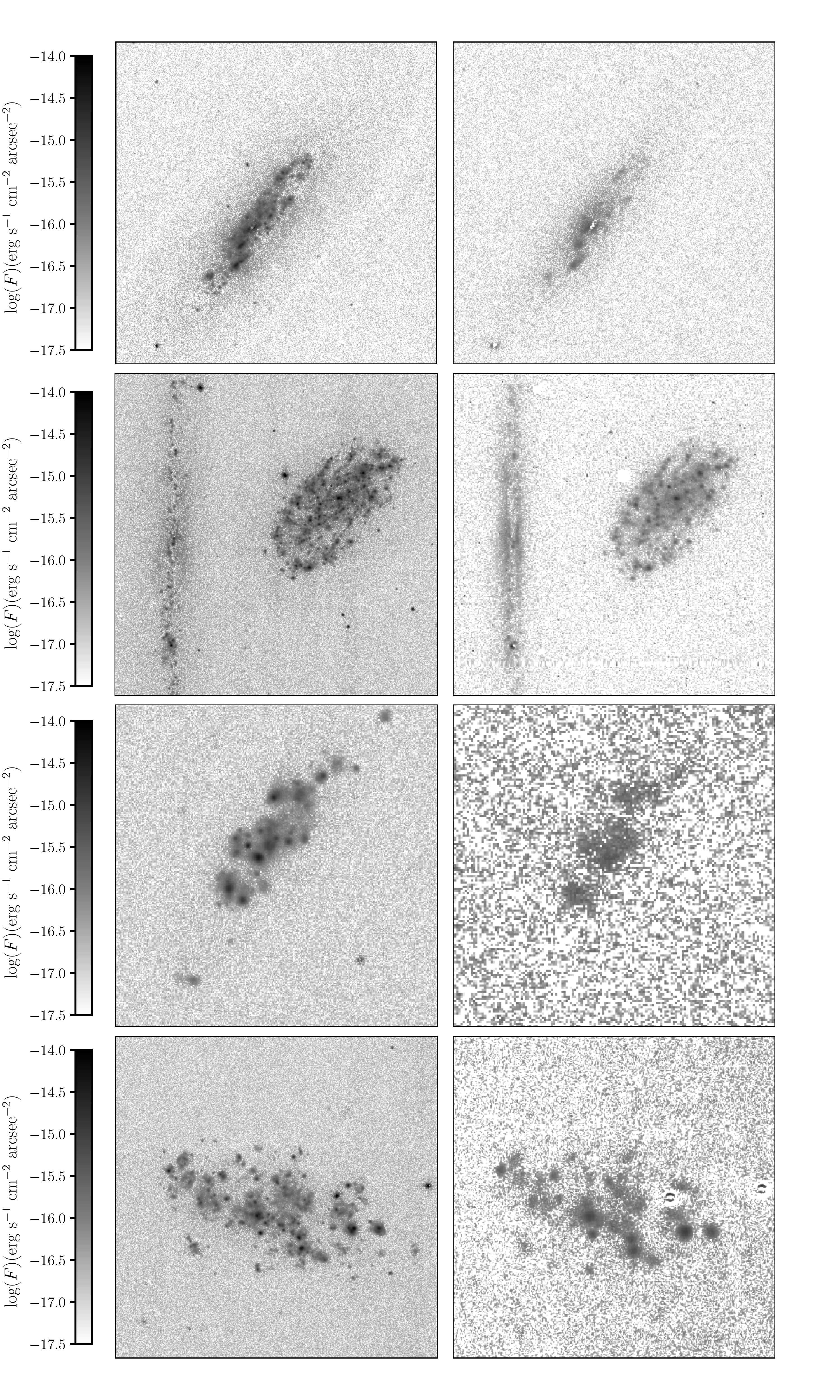}
   \caption{Comparison of the H$\alpha$ images of the galaxies NGC4313 (SA(rs)ab:edge-on), NGC 4298 (SA(rs)c) and NGC4302 (Sc:edge-on), IC3239 (Sm) and IC 3365 (Im) obtained by VESTIGE (left column, from top to
   bottom) to those obtained with a $\sim$ 900 sec on-band exposure on the 2.1 metre telescope at San Pedro Martir (NGC 4313, IC3239, IC3365; Gavazzi et al. 2003b) or a 4000 sec exposure on the 0.9 metre telescope 
   at Kitt Peak (NGC 4298 and 4302; Koopmann et al. 2001) (Right). 
   }
   \label{Comp_Ha}%
   \end{figure}

A huge amount of pointed shallow NB H$\alpha$ imaging data (Young et al. 1996; Macchetto et al. 1996; Koopmann et al. 2001; 
Boselli et al. 2002b, 2015; Boselli \& Gavazzi 2002; Gavazzi et al. 2002a, 
2006; James et al. 2004; Sanchez-Gallego et al. 2012) to which the VESTIGE results can be compared 
already exists in the literature, most of which is available on the GoldMine database (Gavazzi et al. 2003b). 
The sky regions observed during the 2017A campain and the pilot projects include 31 galaxies with published NB H$\alpha$ imaging data.
Fluxes and equivalent widths are compared in Fig. \ref{comp}. The different sets of data are consistent within 5\% ~ for the equivalent widths
and 30\%~ for the fluxes.
The increase in the quality of the NB H$\alpha$ images obtained with MegaCam at the CFHT
with respect to previous targeted H$\alpha$ images is spectacular, as depicted in Fig. \ref{Comp_Ha}. 

\subsection{A 4$\times$1 deg$^2$ strip across the core of the cluster}

   \begin{figure*}
   \centering
   \includegraphics[width=15cm] {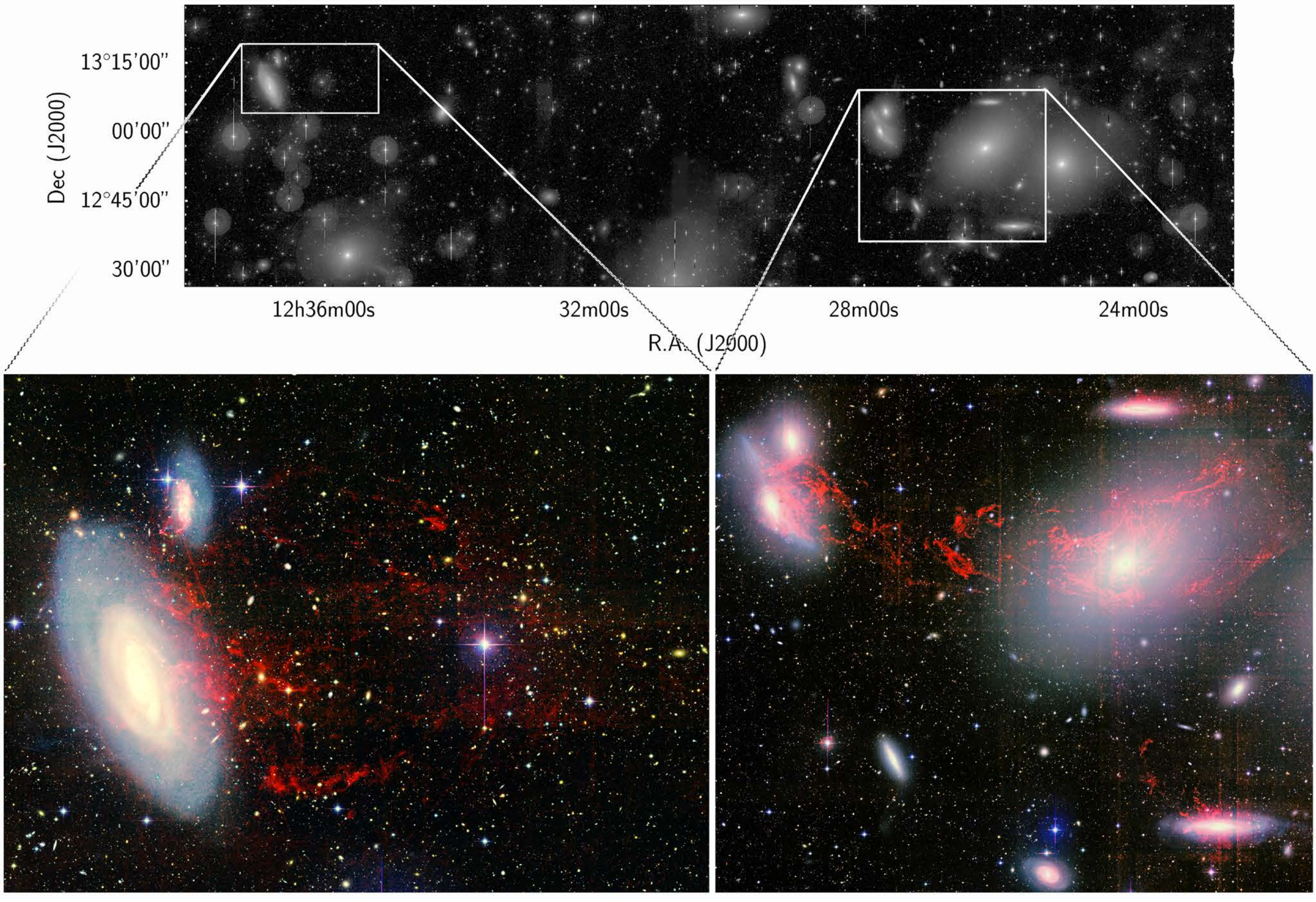} \\
   \caption{Upper panel: $g$-band image of the 4$\times$1 deg$^2$ (corresponding to 1.1$\times$0.3 Mpc$^2$) strip of the core of the cluster north of M87. The lower panels are a magnified 
   view of the boxed regions marked on the upper panel. They show the pseudo-colour images of NGC 4569 and IC 3583 (lower-left panel) and of the NGC 4438-N4388-M86
   complex (lower right panel) obtained combining the NGVS optical $u$ and $g$ in the blue channel, the $r$ and NB in the green, and the $i$ and the 
   continuum-subtracted H$\alpha$ in the red.
 }
   \label{core_rgb}%
   \end{figure*}

   \begin{figure*}
   \centering
   \includegraphics[width=18cm] {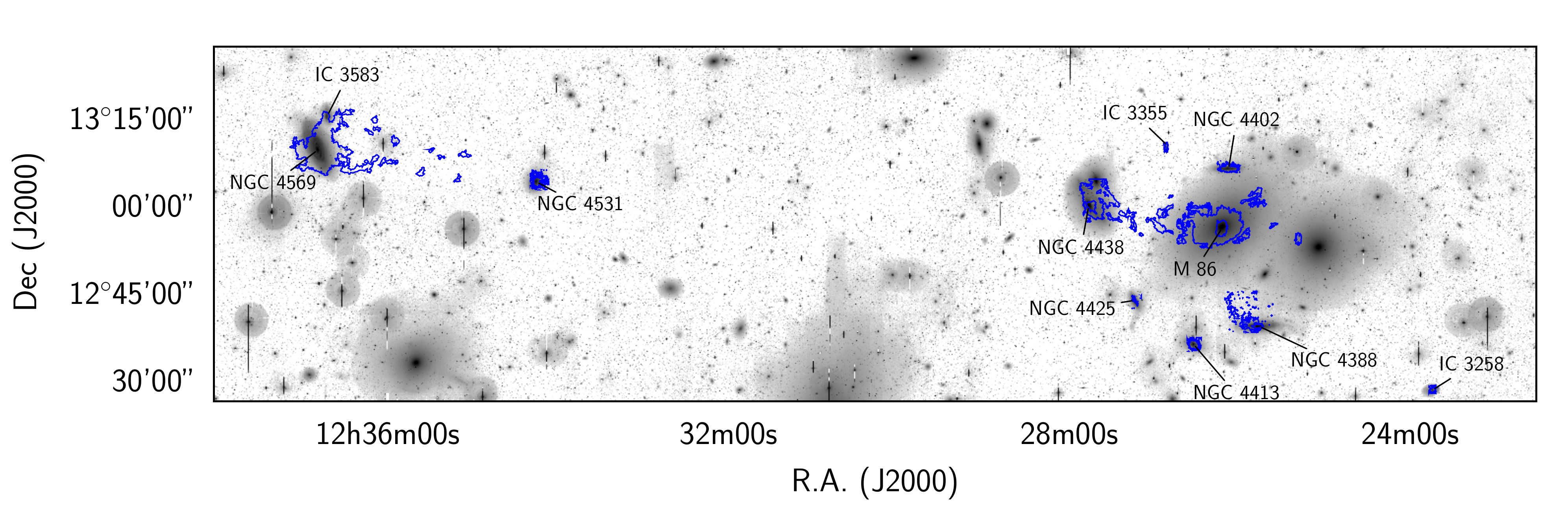} \\
   \caption{Continuum-subtracted H$\alpha$ contours (blue) overplotted on the $r$-band image (grey) of the 
   4$\times$1 deg$^2$ (1.1$\times$0.3 Mpc$^2$) strip of the core of the cluster north of M87. Given that the survey still does not reach 
   its full sensitivity, contours are measured at different surface brightness levels: $\Sigma(H\alpha)$ $\simeq$ 10$^{-18}$ erg s$^{-1}$ cm$^{-2}$ arcsec$^{-2}$ for NGC 4569 and IC 3583
   (from Boselli et al. 2016a),  $\Sigma(H\alpha)$ $\simeq$ 3$\times$10$^{-18}$ erg s$^{-1}$ cm$^{-2}$ arcsec$^{-2}$ for IC 3355, NGC 4388, and NGC 4531, and
   $\Sigma(H\alpha)$ $\simeq$ 6$\times$10$^{-18}$ erg s$^{-1}$ cm$^{-2}$ arcsec$^{-2}$ for M86, IC 3258, NGC 4402, NGC 4413, NGC 4425, and NGC 4438.
 }
   \label{core_cont}%
   \end{figure*}

The observations carried out during the first 2017A semester allowed us to map a large portion of the 
centre of the cluster. Although the current coverage does not yet reach 
the final sensitivity expected for VESTIGE all over the mapped region, these observations already provide the largest NB H$\alpha$ image of the Virgo cluster.
Figure \ref{core_rgb} shows a 4$\times$1 deg$^2$ strip crossing the northern part of the core of the cluster from the giant elliptical galaxy M84 
to NGC 4569, the most massive late-type galaxy of the cluster. Within this strip the survey reaches full sensitivity in the centre and in the eastern
region, while only $\sim$ 20\% in the western part. Figure \ref{core_cont} shows the NGVS $g$-band image of the 
same regions with overplotted H$\alpha$ contours showing emission at low surface brightness levels. 
Several spectacular features are evident in the images: extended filaments of ionised gas are visible to the west 
of NGC 4569 (Boselli et al. 2016a) and to the north-west of NGC 4388 (Yoshida et al. 2002). The image also shows the
spectacular bridge of ionised gas $\simeq$ 215 kpc long (in projected distance) linking M86 to NGC 4438 (Kenney et al. 2008). 
Tails of ionised gas are for the first time detected in NGC 4402 (undetected in the shallower data of Crowl et al. 2005 and Abramson et al. 2016),
NGC 4425, and NGC 4531. The typical surface brightness of these extended features ranges from $\Sigma(H\alpha)$
$\simeq$ 10$^{-18}$ erg s$^{-1}$ cm$^{-2}$ arcsec$^{-2}$ in NGC 4569 (measured on the deeper pilot observations of Boselli et al. 2016a),
to $\Sigma(H\alpha)$ $\simeq$ 6$\times$10$^{-18}$ erg s$^{-1}$ cm$^{-2}$ arcsec$^{-2}$ in the extended filaments of NGC4438 and M86, 
where the sensitivity of the survey is lower. 
This image witnesses a large variety of extended ionised gas morphologies, 
possibly caused by different hydrodynamical processes (tidal interactions in NGC 4438 and M86, ram pressure in NGC 4388, NGC 4402, NGC 4425, NGC 4531, and NGC 4569) 
proving that VESTIGE is a revolutionary survey in probing environmental effects in the Virgo cluster. 

\section{Scientific objectives}

\subsection{Virgo cluster science}

\subsubsection{The effects of the environment on galaxy evolution}

We expect to detect low surface brightness ionised gas features extending across several tens of kpc in several 
late-type galaxies in the cluster and its surrounding regions. These features result from ongoing stripping processes, removing 
the gaseous content of star-forming galaxies as they enter the cluster. They can also occur in massive elliptical galaxies formed 
by major merging events (M87, Gavazzi et al. 2000; M86, Kenney et al. 2008), and in early-type dwarfs formed by quenching of 
low-mass star-forming systems (Boselli et al. 2008a). These processes are thought to play a key role 
in the transformation of star-forming galaxies into quiescent systems, and in the build-up of the red sequence (Balogh et al. 2000; Boselli \& Gavazzi 2014). 
Identifying the exact nature (i.e., ram pressure stripping, harassment, starvation, tidal stripping) of the perturbing mechanisms, 
as a function of environment (from the cluster core to its periphery), is crucial for constraining cosmological models of galaxy 
evolution. Meanwhile, gravitational perturbations such as tidal stirring and harassment are expected to perturb simultaneously the 
gaseous and stellar components, producing low surface brightness tails in both the H$\alpha$ and in the optical broad-band images. 
Gravitational perturbations can be identified by comparing visual signs of interactions with
quantitative measurements (e.g., asymmetires - Conselice 2003; Gini coefficient - Abraham et al. 2003; $M_{20}$ - Lotz et al. 2004)
and with the distance from the closest companion (Patton et al. 2016).
The dynamical interaction of the galaxy ISM with the hot ICM, on the contrary, should affect only the gaseous component (e.g. Fumagalli et al. 2014). 
The VESTIGE data will be used to: (1) determine the fraction of galaxies with signs of perturbation due to gravitational interactions 
or interactions with the ICM (Yagi et al. 2010); (2) identify and map in the cluster the transition dwarf galaxies, which are believed to be in the midst 
of a transformation from dwarf irregular to early-type dwarfs (Boselli et al. 2008a, 2008b, 2014a; Cot\'e et al. 2009; De Looze et al. 2013); 
(3) determine the physical extent, the density, and total mass of the 
gas in the ionised phase (making simple assumptions on the filling factor) and compare it to that of the other gas phases (Fossati et al. 2016; Boselli et al. 2016a); (4) identify 
extraplanar HII regions, possible progenitors of intracluster star clusters, that can contribute to the ionisation of the gas
(Fumagalli et al. 2011b; Arrigoni Battaia et al. 2012; Fossati et al. 2016; Boselli et al. 2017); and (5) 
compare the observations to the predictions of published (Tonnesen \& Bryan 2010, 2012; Tonnesen et al. 2011; Tonnesen \& Stone 2014) or upcoming hydrodynamic simulations of gas 
stripping developed by our team and/or presented in state-of-the-art cosmological simulations (Genel et al. 2014; Schaye et al. 2015; Barnes et al. 2017), 
as illustrated in Fig. \ref{simulazioni}.

\subsubsection{The fate of the stripped gas in cluster galaxies}

The different phases of the ISM 
(dust, neutral atomic and molecular gas, ionised gas, hot gas) within galaxies can be removed during interactions with the harsh cluster 
environment. The unprecedented multifrequency data available for Virgo, combined with tuned models of gas stripping and follow-up 
spectroscopic observations that we plan to obtain with SITELLE (Drissen et al. 2010) and MUSE (as in, e.g., Fumagalli et al. 2014, Fossati et al. 
2016, Consolandi et al. 2017b), will be used to study the fate of this stripped material in the cluster environment. The main H$\alpha$ streams will be searched for 
in atomic, molecular  and X-ray gas, to test the survival of cold gas in the cluster hot gas environment (Serra et al. 2013; Jachym et al. 2013, 2014; Verdugo 
et al. 2015). The data will be used to characterise the physical conditions (metallicity, density, temperature, turbulence) under which 
the stripped gas in its different phases (neutral, ionised, hot) can collapse to form new stars. These data are essential for constraining 
hydrodynamic simulations of gas stripping (Tonnesen \& Bryan 2010, 2012). Simulations tailored to VESTIGE will be 
carried out with the 3D hydrodynamic Adaptive Mesh Refinement code FLASH (Fryxell et al. 2000) and Enzo (Bryan et al. 2014). FLASH will be updated to include a 
multi-phase ISM and the stellar component (Mitchell et al. 2013), as well as the thermal conduction by the hot ICM. 
The treatment of the different gas phases and their energetic interactions is crucial for a complete and coherent understanding of the 
physical process and, in turn, for constraining simulations with observational data. Systematic and complete surveys in H$\alpha$, HI, CO and 
X-rays are essential if we are to sample with high statistical significance the largest possible range in the ICM, ISM and galaxy parameter 
space.

\subsubsection{The star forming process in nearby galaxies}

Within galaxies, the H$\alpha$ emission line originates primarily from gas ionised by young and massive OB stars; 
it is thus an excellent tracer of ongoing star formation (Kennicutt 1998; Boselli et al. 2001, 2009; Kennicutt \& Evans 2012),
sensitive to stellar populations much younger than those detected in the GALEX UV bands (H$\alpha$ $\lesssim$ 10 Myr vs. UV $\lesssim$ 100 Myr). 
VESTIGE will provide the largest and 
most homogeneous database ever assembled with which to study star formation down to sub-kpc scales, both in normal and perturbed galaxies 
and along the luminosity function, all the way down to the faintest blue compact dwarfs/HII galaxies ($L(H\alpha)$ $\sim$ 10$^{36}$ 
erg s$^{-1}$, equivalently to $SFR$ $\sim$ 10$^{-5}$ M$_{\odot}$ yr$^{-1}$ when the $SFR$ is derived using the standard calibration of Kennicutt 1998) and in virtually all star-forming systems with 
$M_{star}$ $\sim$ 10$^5$ M$_{\odot}$.
At these SFRs, we will be able to study the stochastic 
sampling of the IMF (Fumagalli et al. 2011a) and assess its impact on the calibration of the H$\alpha$ luminosity as a star formation tracer 
(Boselli et al. 2009; da Silva et al. 2014). Dust attenuation will be estimated by comparing the H$\alpha$ emission to the emission of dust in the mid- and 
far-infrared following standard recipes (Calzetti et al. 2010; Kennicutt et al. 2009; Zhu et al. 2008;
Boselli et al. 2015) or using spatially resolved spectral energy distribution 
fitting analysis based on specific codes developed within our team (CIGALE, Boquien et al. 2012, 2014, 2016; Boselli et al. 2016b).  The contribution 
of the [NII] lines 
can be determined using the long-slit data collected so far for the bright galaxies (Gavazzi et al. 2004; Boselli et al. 2013) or from known scaling 
relations (Decarli et al. 2007; Boselli et al. 2009), 
while possible AGN contamination can be quantified by nuclear spectroscopy available for 86\% of the galaxies of the sample with $g$ $<$ 17.5 
mag ($\simeq$ 1000 objects; Decarli et al. 2007; Gavazzi et al. 2013; SDSS). Additionally, we will be able to address the interplay between starbursts and AGN activity on a representative
sample of $\sim$ 1000 galaxies. The observed 2D properties of the star-forming activity within galaxies will also be compared to the molecular and 
atomic gas column densities (Boissier et al. 2001, 2003a, 2003b, 2007, 2008) and to the predictions of multi-zone chemo-spectrophotometric models of galaxy evolution 
specifically tailored to take into account the effects of the environment (Boselli et al. 2006, 2008a, 2008b, 2014a). 

The imprint of secular evolution on the star formation history of galaxies will be searched for by analysing the relationship between 
the presence of bars in the broad-band stellar images and in the NB H$\alpha$ images, extending the work of Gavazzi et al. (2015),
Consolandi (2016), and Consolandi et al. (2016, 2017a) to low surface brightness dwarfs unreachable by the SDSS because of its limited sensitivity and 
angular resolution. The data will also be compared to the predictions of chemodynamical simulations of galaxy evolution 
(Michel-Dansac \& Wozniak 2004). The comparison between this unique set of multifrequency imaging and spectroscopic data for hundreds 
of resolved galaxies will be a powerful tool for reconstructing the 2D star formation history of galaxies and studying the quenching 
process as a function of mass, morphological type and environment.

\subsubsection{The ionised gas emission in early-type galaxies}

\begin{figure*}
   \centering
   \includegraphics[width=18cm]{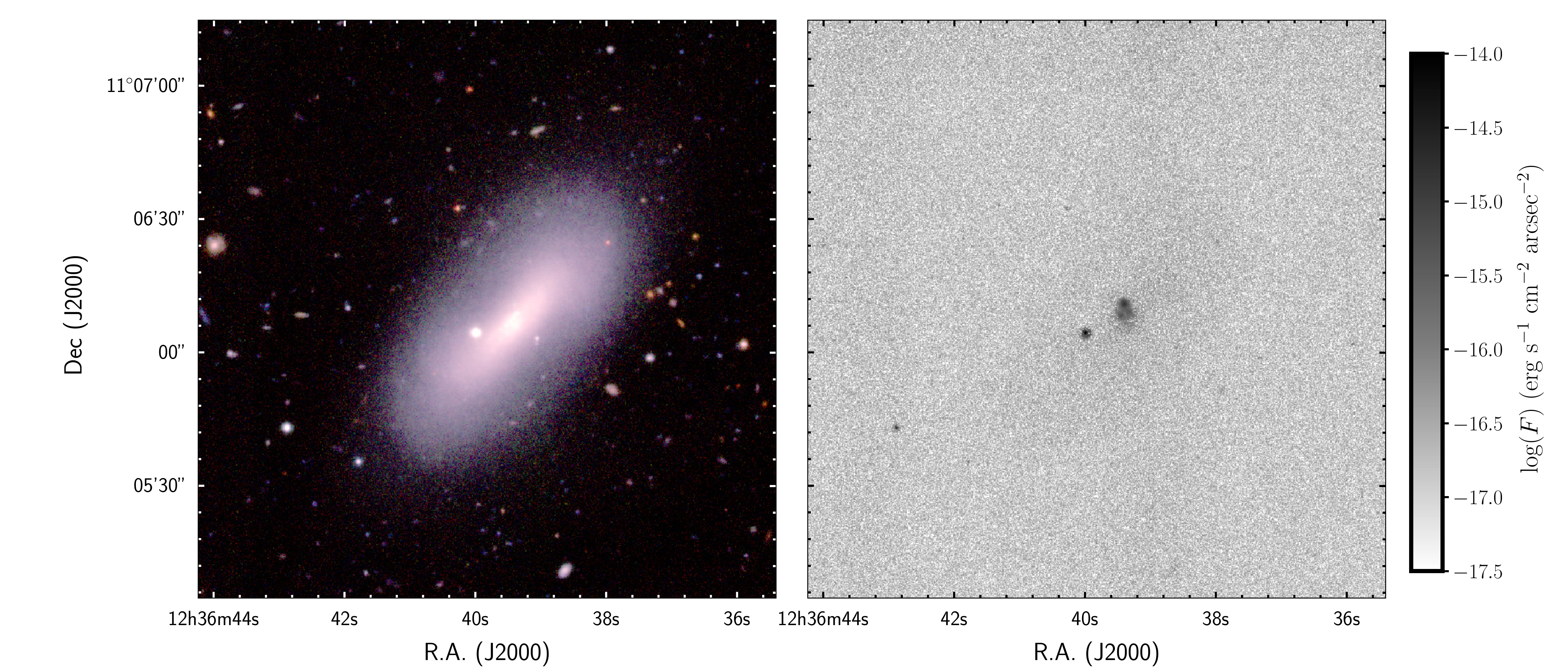}
   \caption{The NGVS $ugi$ rgb colour image (left) and the VESTIGE continuum-subtracted H$\alpha$ image of the dwarf spheroidal galaxy (dS0(8):; Binggeli et al. 1985) IC 3578 (VCC 1684). The H$\alpha$
   image shows the presence of a few compact HII regions in the nucleus of the galaxy, witnessing an ongoing nuclear star forming activity.
 }
   \label{dE}%
   \end{figure*}

\begin{figure*}
   \centering
   \includegraphics[width=18cm]{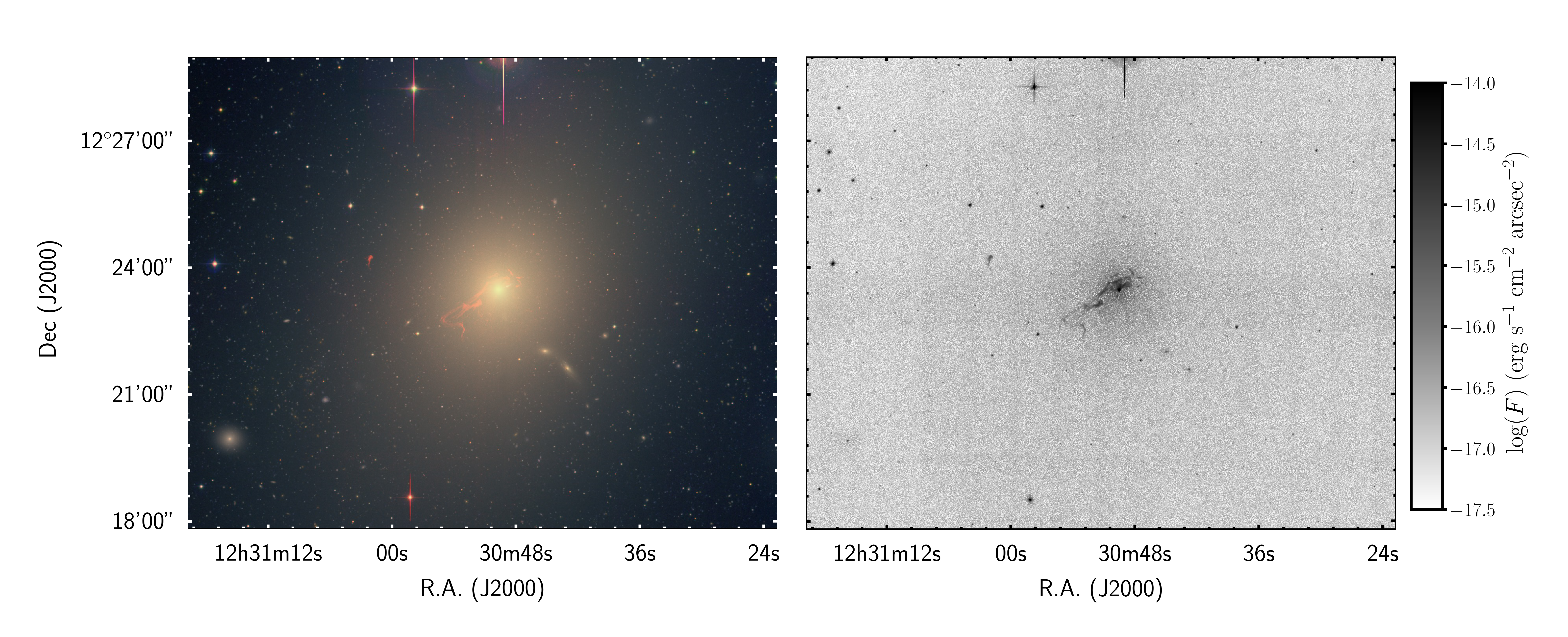}
   \caption{Left: pseudo-colour image of M87 obtained combining the NGVS optical $u$ and $g$ in the blue chanel, the $r$ and NB in the green, and the $i$ and the 
   continuum-subtracted H$\alpha$ in the red. Right: continuum-subtracted H$\alpha$ emission of M87. Both images show the presence of prominent filaments of ionised gas 
   extending $\sim$ 10 kpc out from the nucleus. 
 }
   \label{M87}%
   \end{figure*}

Low levels of H$\alpha$ emission will also be 
detected in massive early-type galaxies (Gomes et al. 2016; Belfiore et al. 2016; Gavazzi et al. 2017), while residual star formation is expected in the core of the 
recently formed dE galaxies (Boselli et al. 2008a; see Fig. \ref{dE}), which are often characterised by young nuclear star clusters or cores (C\^ot\'e et al. 2006, Lisker 
et al. 2006). 
Given the sensitivity of the survey in terms of surface brightness, 
we also expect to detect a weak diffuse emission originating from evolved stellar populations in dwarf ellipticals (Michielsen et al. 2004), 
or filaments of ionised gas in massive galaxies such as those observed in M87 (Sparks et al. 1993; 
Gavazzi et al. 2000; Fig. \ref{M87}) and M86 (Trinchieri \& di Serego Alighieri 1991; Kenney et al. 2008).
Within the VESTIGE footprint there are 46 elliptical galaxies, 68 lenticulars, 13 S0a, and 1112 dE/dS0 galaxies catalogued in the Virgo Cluster 
Catalogue (Binggeli et al. 1985) with a recessional velocity $vel$ $<$ 3500 km s$^{-1}$, 
while another few thousands dwarfs, classified as Virgo members, have been detected by the NGVS survey (Ferrarese et al. 2012, 2016).
VESTIGE will thus provide a unique view of the ionised gas emission properties of early-type galaxies in a complete and statistically significant sample.

\subsubsection{The H$\alpha$ luminosity function}

\begin{figure}
   \centering
   \includegraphics[width=9cm]{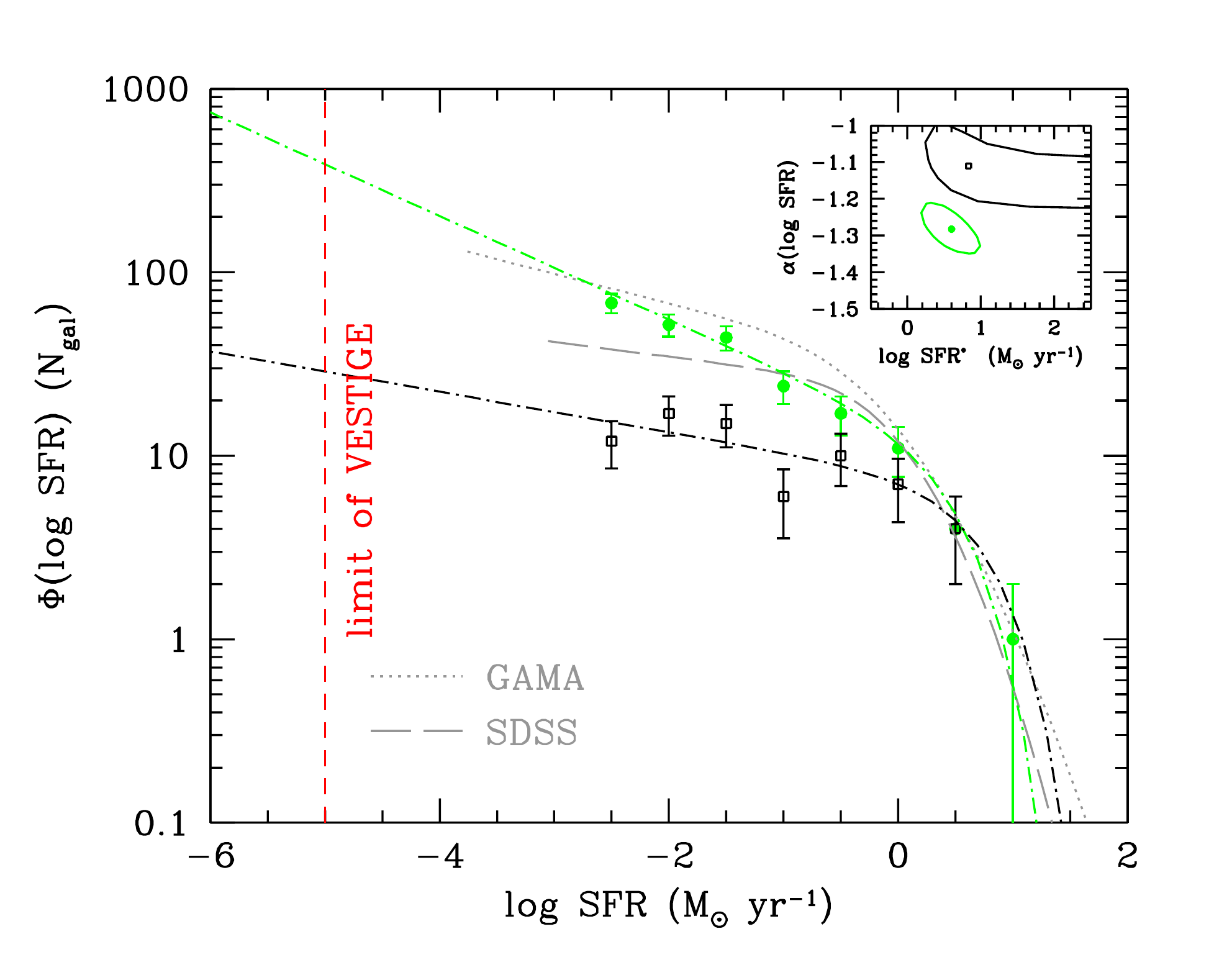}
   \caption{The extrapolated star formation rate (SFR) luminosity function of star-forming galaxies in the Virgo cluster
   periphery (green filled dots, green dot-dashed line) and in subcluster A (black open squares, black dot-dashed line) derived from the GUViCS NUV
   luminosity function (Boselli et al. 2016c) are compared to those derived for field galaxies by Gunawardhana et al. (2013) for the GAMA (grey
   dotted line) and SDSS (grey long dashed line) samples. VESTIGE will extend by two orders of magnitude any other SFR luminosity function available
   in the literature, sampling SFR as low as $\simeq$ 10$^{-5}$ M$_{\odot}$ yr$^{-1}$. The estimated total number of star-forming systems
   VESTIGE will detect will be $\sim$ 500. 
 }
   \label{LF}%
   \end{figure}

The limit in H$\alpha$ luminosity or in star formation rate $SFR$ that VESTIGE will reach is $\sim$ 100 times fainter than the limits reached 
by the most recent local H$\alpha$ luminosity functions (Gunawardhana et al. 2013; Fig. \ref{LF}). 
By counting directly the number of star-forming objects in the NGVS footprints down to $M_{star}$ $\sim$ 10$^5$ M$_{\odot}$ 
we expect to detect $\sim$ 500 objects. This roughly corresponds to the number derived  
by extrapolating the SFR luminosity function derived from the UV data (Boselli 
et al. 2016c) to $SFR$ = 10$^{-5}$ M$_{\odot}$ yr$^{-1}$, and to the number of blue galaxies detected by NGVS.
This number is perfectly suited for an accurate determination of the best fitting parameters in a parametric luminosity function both within the cluster core
or in the cluster periphery.

\subsubsection{The H$\alpha$ scaling relations}

\begin{figure}
   \centering
   \includegraphics[width=9cm]{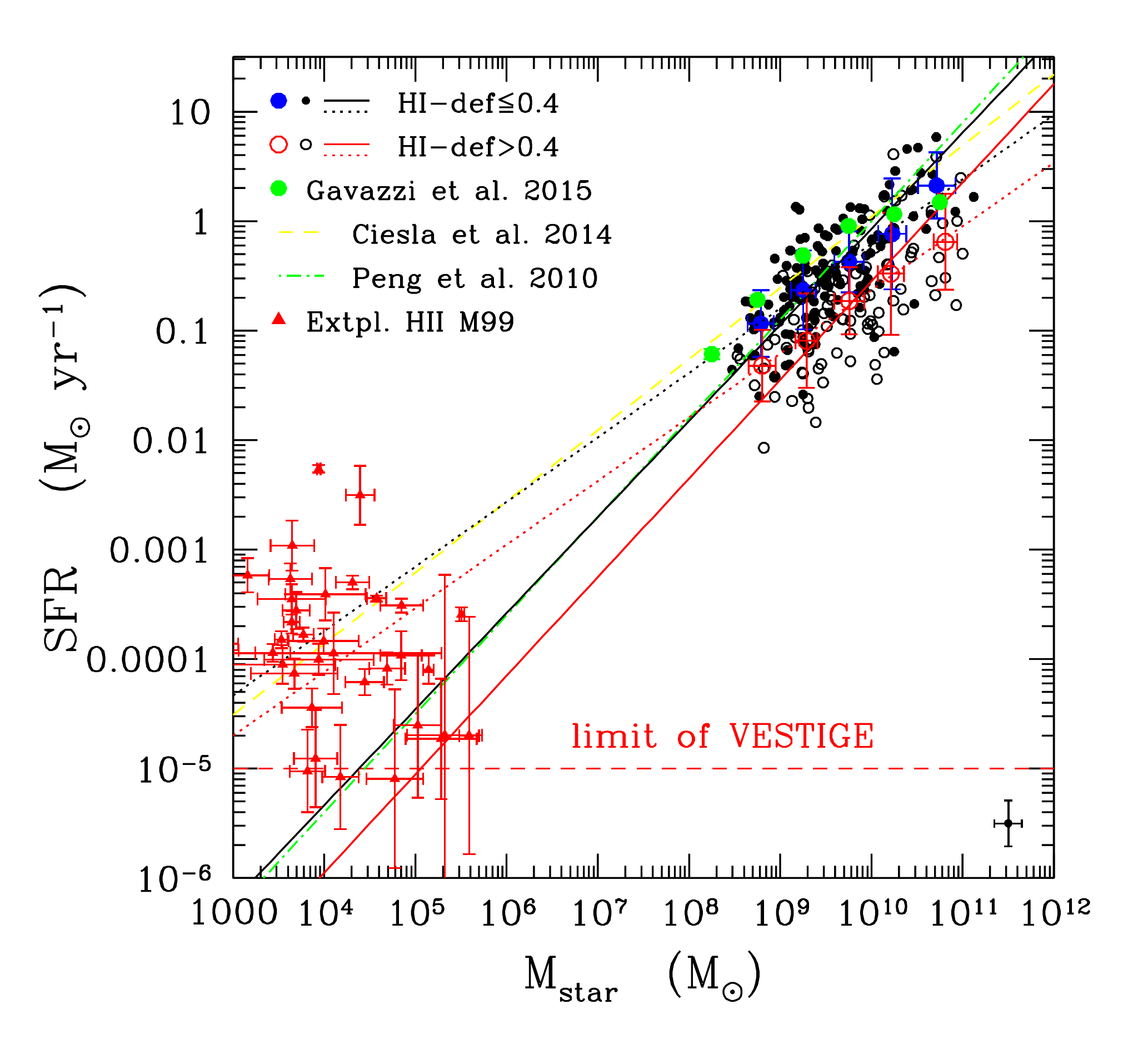}
   \caption{The relationship between the star formation rate and the stellar mass (main sequence) derived for the \textit{Herschel} Reference Survey (HRS) late-type galaxies by Boselli et al. (2015).
   Filled dots are for HI-normal field galaxies, empty symbols for HI-deficient Virgo cluster objects. The large filled blue and empty red circles give the mean values in different bins of
   stellar mass. The large filled dots indicate the mean values of Gavazzi et al. (2015). The solid black and red lines are the bisector fit for 
   HI-normal and HI-deficient galaxies, while the dotted lines are the linear fits. The linear best fit of Peng et al. (2010) is shown by the green dotted-dashed line, while that 
   of Ciesla et al. (2014) by the yellow dashed line. The error bar in the lower right corner shows the typical uncertainty on the data (adapted from Boselli et al. 2015).
   The detection limit of VESTIGE will be $\simeq$ 10$^{-5}$ M$_{\odot}$ yr$^{-1}$. Red triangles indicate the extraplanar HII regions of NGC 4254 (M99)
   detected by VESTIGE and formed after a gravitational perturbation with a nearby object (Boselli et al. paper IV). 
   }
   \label{main}%
   \end{figure}

As for the determination of the H$\alpha$ luminosity function, VESTIGE will provide us with the best sample of galaxies with H$\alpha$ data suitable
for the determination of the H$\alpha$ and $SFR$ scaling relations down to the dwarf galaxy regime, significantly increasing in terms of sensitivity
and statistics the most recent studies (Gavazzi et al. 2013; Boselli et al. 2015). The data will be used, for instance, to extend by three orders of magnitude 
the most recent local determinations of the star formation main sequence (Fig. \ref{main}). These scaling relations will be determined at different
clustercentric distances, from the densest regions in the core of cluster A, to the cluster periphery. Any possible observed clustercentric variation 
in the typical H$\alpha$ scaling relations, for the first time determined on a strong statistical basis, will be compared to the predictions of cosmological simulations and
hydrodynamical models of galaxy evolution for the identification of the dominant perturbing process in rich environments. 

\subsubsection{The nature of almost dark galaxies}

Extended HI sources without any stellar counterpart have been detected in the Virgo cluster by AGES (Taylor et al. 2012) and ALFALFA (Janowiecki et al.
2015; Cannon et al. 2015). These "dark galaxies" are particularly important since they have been proposed as a possible solution to the missing
satellite problem (Bullock 2010). Their nature, however, is still poorly known. In the most massive clouds ($\sim$ 10$^9$ M$_{\odot}$),
the atomic gas should form molecular hydrogen, become unstable, collapse and form new stars (Taylor \& Webster 2005; Burkhart \& Loeb 2016). 
In lower mass objects, on the contrary, the contact with the hot intracluster medium would make the gas cloud change phase
to reach the typical temperature ($T$ $\sim$ 10$^7$-10$^8$ K) of the surrounding gas  (thermal evaporation, Cowie \& Songaila 1977).
VESTIGE might thus detect low levels of star formation within massive HI-selected dark galaxies (see Fig. \ref{HVC}) as well as the extended emission associated
to any possible phase transition within the clouds. It can also discover new dark galaxy candidates as extended H$\alpha$ blobs 
without any visible stellar counterpart, for which follow-up spectroscopic observations will be required to confirm their extragalactic nature
and characterise the physical properties of the ionised gas.

\subsubsection{The dynamical structure of the Virgo cluster}

The spectacular tails of ionised gas detected in H$\alpha$ (that can extend up to $\sim$ 100 kpc) trace the trajectory on the plane of the sky of 
galaxies that have recently entered the cluster environment (Fig. \ref{simulazioni}). Combined with radial velocities, which are available for more than 1000 objects 
within the selected region and velocity range (-1000 $<$ $cz$ $<$ 3500 km s$^{-1}$), and specific numerical dynamical codes (e.g. Bovy 2015)
the orientation of the extended tails of ionised gas will be 
used to reconstruct the orbits of individual galaxies (Vollmer 2009). These data will allow the first-ever study of the dynamical evolution of 
the cluster in three dimensions. Characterising the dynamical properties of the cluster is essential for a fair comparison with cosmological 
simulations (Vollmer et al. 2001) and, along with the physical properties of galaxies, will be of prime importance for understanding 
the role of pre-processing in galaxy evolution. 

\subsubsection{The HII region luminosity function of cluster galaxies}

The sensitivity of the survey for point sources and the excellent image quality will allow us to measure the first HII region luminosity function of cluster galaxies (Kennicutt 1981).
VESTIGE will be able to detect classical HII regions of luminosity $L(H\alpha)$ $\gtrsim$ 10$^{36}$ erg s$^{-1}$. Considering a typical seeing of 0.7 arcsec,  
corresponding to $\simeq$ 60 pc at the distance of Virgo, VESTIGE will be able to resolve only giant ($L(H\alpha)$ = 10$^{37}$-10$^{39}$ erg s$^{-1}$)
and super giant ($L(H\alpha)$ $\simeq$ 10$^{39}$ erg s$^{-1}$) HII regions. Although still limited in terms of sensitivity and angular resolution with respect to dedicated HST studies of 
nearby galaxies (Scoville et al. 2001; Lee et al. 2011), or galaxies in the Local Group 
(Kennicutt \& Hodge 1986; Hodge et al. 1990, 1999), the improvement with respect to other studies based on ground-based images
(Kennicutt et al. 1989; Rand 1992; Youngblood \& Hunter 1999; Thilker et al. 2002; Bradley et al. 2006)
will be significant in terms of sampled range of H$\alpha$ luminosity, angular resolution, and statistics.

\subsubsection{Planetary nebulae and the origin of the intracluster light}

Even though planetary pebula (PN) surveys in distant galaxies focus on the [OIII] $\lambda$ 5007 \AA\ emission line due to its strength, 
a PN spectrum is composed by several other emission lines of which H$\alpha$ is one of the strongest (Ciardullo 2010). 
H$\alpha$ studies of PN systems have been limited to the local Universe and very little is known about the 
PN properties at this wavelength for more distant objects like Virgo cluster galaxies. Ciardullo et al. (2010) presented the 
H$\alpha$ luminosity function (LF) for three Local Group galaxies showing that like the [OIII] PNLF, the H$\alpha$ LF displays 
a cutoff that is insensitive to stellar population. The same cutoff is also visible in the recent SITELLE data of M31 (Martin et al. 2018). 
At the distance of the Virgo cluster, this cutoff
should be at $f(H\alpha)$ $\simeq$ 3 $\times$ 10$^{-17}$ erg s$^{-1}$ cm$^{-2}$. This number is very close to the detection limit
of the survey ($f(H\alpha)$ $\simeq$ 4 $\times$ 10$^{-17}$ erg s$^{-1}$ cm$^{-2}$ 90\% complete at 5 $\sigma$, 
$f(H\alpha)$ $\simeq$ 2.5 $\times$ 10$^{-17}$ erg s$^{-1}$ cm$^{-2}$ 50\% complete at 5 $\sigma$). However, given the sharp increase of the PNLF
at its bright end, and the large scatter in the [OIII]/H$\alpha$ ratio observed in nearby galaxies, VESTIGE can potentially detect 
the brightest PNe in several tens of galaxies over the 104 deg$^2$ covered by the survey.

\begin{figure}
   \centering
   \includegraphics[width=8cm]{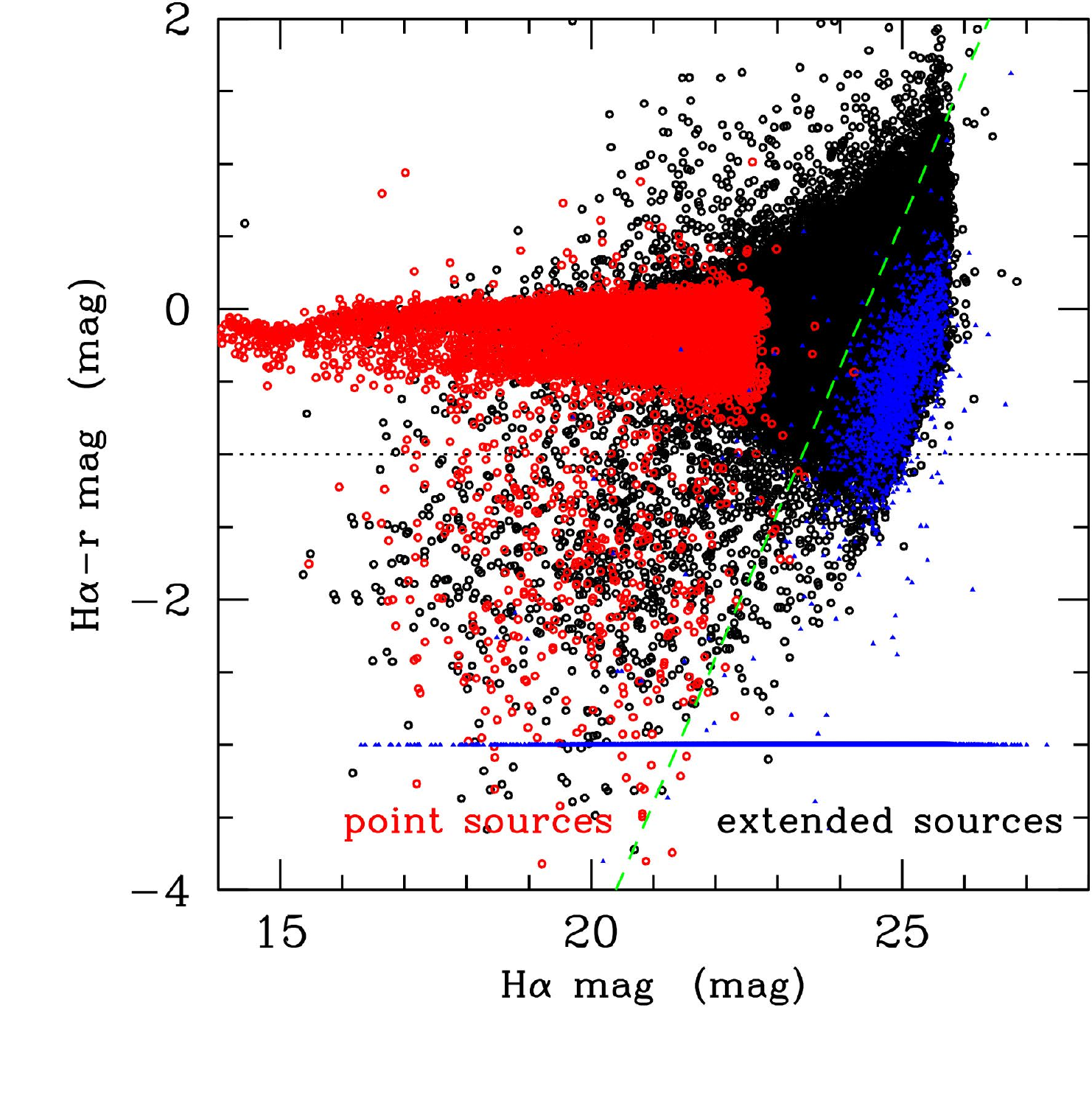}
   \caption{The H$\alpha$-$r$ vs. H$\alpha$ colour magnitude diagram for sources in the field of NGC 4302. Red symbols 
   are point sources, black symbols extended sources, while blue symbols show
   those objects detected in the NB H$\alpha$ image with a $r$-band detection of S/N$\leq$ 5 (those with H$\alpha$-$r$ $>$ -3) or without any counterpart in the $r$-band ( H$\alpha$-$r$ $=$ -3). 
   The colour-dependent detection limit of VESTIGE is indicated by the green dashed line. Those sources below the colour magnitude relation,
   with a colour H$\alpha$-$r$ $\leq$ -1 mag (black dotted line)
   are line emitters and include local galaxies, PNe, and background line ([OIII], [OII], and Ly$\alpha$) emitting galaxies. }
   \label{pointsource}%
   \end{figure}

PNe will be identified using standard (H$\alpha$-$r$) vs. H$\alpha$ colour diagrams such as the one plotted in Fig. \ref{pointsource} 
and distinguished from high-$z$ line emitters or local HII regions from the lack of any stellar emission in the deep $g$ and $i$ NGVS images 
(Jacoby et al. 1990; Theuns \& Warren 1997; Ciardullo et al. 1998; Mendez et al. 2001; see however Bacon et al. 2017). 


\subsection{Foreground science}

\subsubsection{The diffuse ionised medium of the Milky Way}

\begin{figure}
   \centering
   \includegraphics[width=9cm]{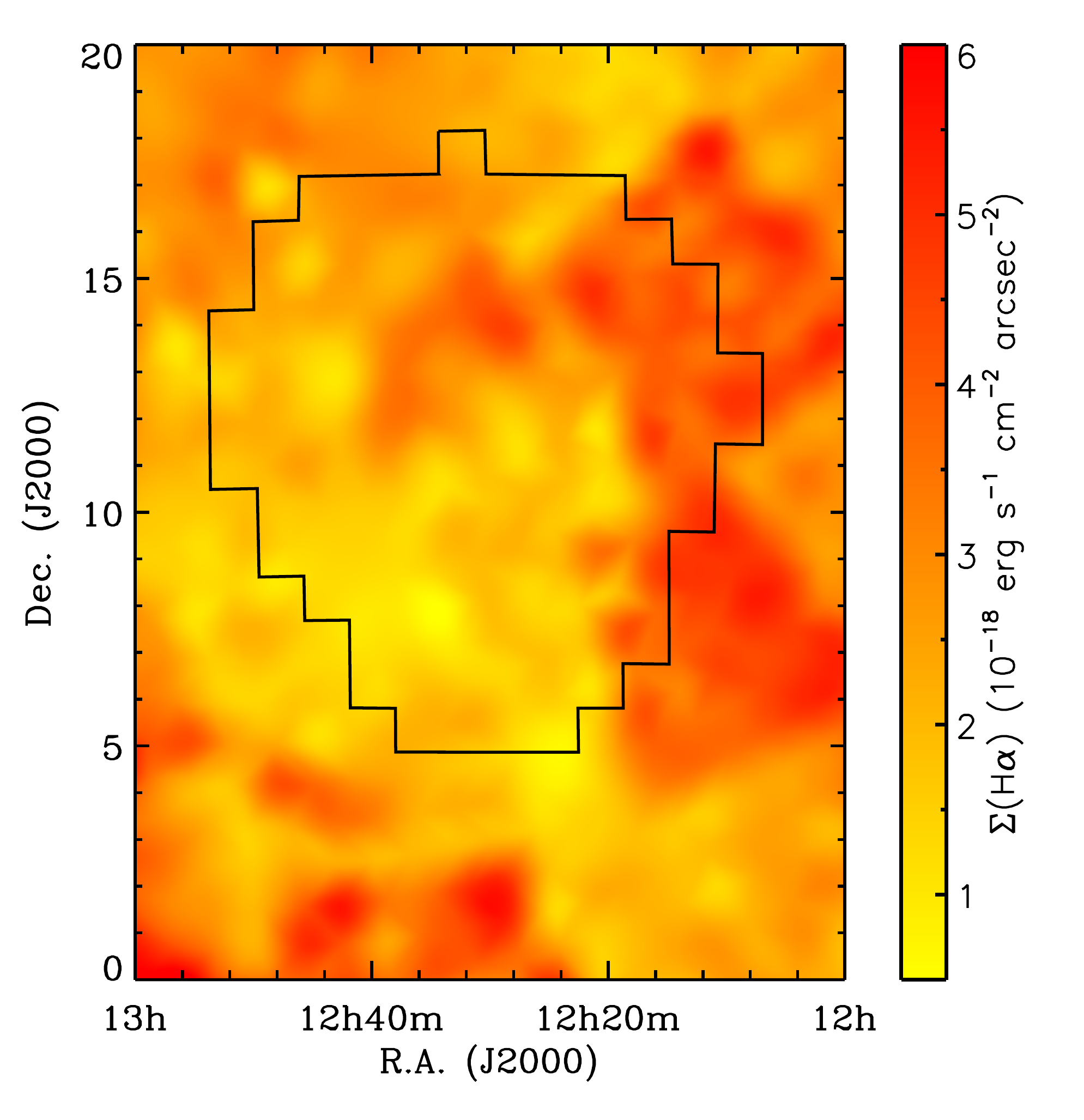}
   \caption{The diffuse H$\alpha$ emission of the Milky Way in the direction of the Virgo cluster at an angular resolution of 1 deg, 
   as derived from the WHAM all-sky survey (Reynolds et al. 1998). 
 }
   \label{MW}%
   \end{figure}

The H$\alpha$ filter is sensitive to the diffuse emission from ionised gas in the Milky Way (e.g. Reynolds et al. 1998). At the 
high Galactic latitude of Virgo ($b$ $\sim$ 74$^o$), the diffuse Galactic H$\alpha$ emission is $\sim$ 0.1-1.0 Rayleigh (corresponding to 
$\Sigma (H\alpha )$ $\sim$ 0.5-5 $\times$ 10$^{-18}$ erg sec$^{-1}$ cm$^{-2}$ arcsec$^{-2}$) as derived by the Wisconsin H$\alpha$ Mapper (WHAM) 
all-sky survey (Reynolds et al. 1998, Fig. \ref{MW}) and can 
thus be detected once the signal is appropriately smoothed. The Milky Way emission can be distinguished from large-scale fake structures in the 
continuum-subtracted images such as those produced by poor quality flat-fielding or bright star halos, or from the tails of ionised gas 
associated with Virgo cluster galaxies, once the full mosaic of images is in hand. The H$\alpha$ emission of the Galactic halo, indeed, 
is expected to extend over several square degrees, leading to the formation of long and continuous tails covering several frames. 
For this purpose we are developing within the team a modified version of the Elixir-LSB data reduction pipeline which uses the
1$^o$ resolution WHAM map (Fig. \ref{MW}) as a prior to fix a zero point for the diffuse Galactic emission over each MegaCam frame.HI and 
CO line widths will be used to discriminate Galactic from extragalactic emission (Cortese et al. 2010b; For et al. 2012), a technique we already have used 
successfully to identify and map the distribution of the scattered light produced by the Galaxy cirrus emission in GALEX far- and near-UV 
images of the same Virgo cluster region (Boissier et al. 2015) or that emitted by the Galactic cirrus in the far-infrared 
(Bianchi et al. 2017). VESTIGE will provide a substantial improvement in angular resolution 
with respect to WHAM, producing the first H$\alpha$ map of the Milky Way at high Galactic latitude over a 
$\sim$ 100 deg$^2$ contiguous field and at a few 
arcsec resolution. It will also be complementary to shallower ($\Sigma (H\alpha)$ $\sim$ 2$\times$ 10$^{-17}$ erg sec$^{-1}$ cm$^{-2}$ arcsec$^{-2}$) 
surveys limited to the Galactic 
plane (IPHAS, Drew et al. 2005; VPHAS, Drew et al. 2014; SHASSA, Gaustad et al. 2001; VTSS, Dennison et al. 1998; SHS, Parker et al. 2005). 
The H$\alpha$ data will be compared to those available for other components of the ISM (i.e., dust from IRAS, \textit{Planck} and \textit{Herschel}; HI gas from 
GALFA, Peek et al. 2011 and ALFALFA, Bianchi et al. 2017;scattered light from GALEX, Boissier et al. 2015) to firmly characterise the physical properties of the Milky Way's ISM at high 
Galactic latitudes (Boulanger et al. 1996; Lagache et al. 2000).

\subsubsection{High velocity clouds, compact sources and Galactic fountains}

\begin{figure}
   \centering
   \includegraphics[width=10cm]{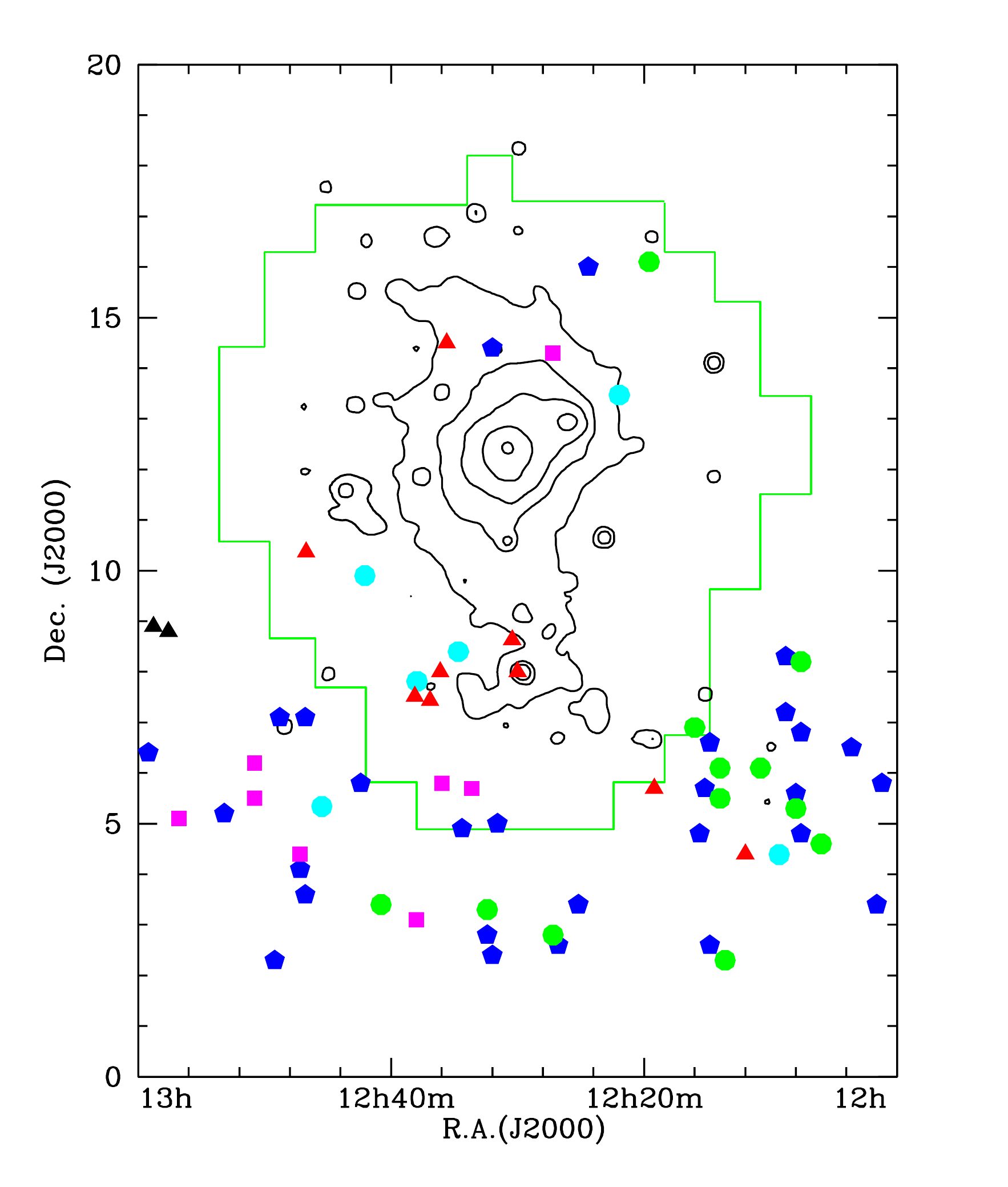}
   \caption{Distribution of the HI compact sources detected by GALFA-HI and ALFALFA within the Virgo cluster region mapped by VESTIGE. The green footprint indicates the 
   blind mapped region of the survey.  
   The black contours show the X-ray emitting hot gas distribution within the cluster. Black filled triangles are HVC, 
   red filled triangles galaxy candidates and dark galaxies (see Sect. 7.1.7), blue filled pentagons cold LVC, magenta filled squares warm LVC, green filled dots warm positive LVC (from Saul et al. 2012), 
   while cyan filled dots UCHVC, from Adams et al. (2013).}
   \label{HVC}%
   \end{figure}

The HI Galactic survey done with Arecibo (GALFA-HI, Saul et al. 2012) detected 13 compact sources within the VESTIGE footprint (see Fig. \ref{HVC}),
and others will be detected by Wallaby (Koribalski 2012). 
These sources have been identified as high velocity clouds (HVC), galaxy candidates, cold and warm low velocity clouds (LVC). A different population of
ultra-compact high velocity clouds (UCHVC) has been also detected by ALFALFA, probably associated to very low mass galaxies 
in the Local Volume (Adams et al. 2013; Bellazzini et al. 2015). Four of these are within the VESTIGE footprint. 
Given their low recessional velocity ($\leq$ 400 km s$^{-1}$), all these 
sources are potentially detectable by VESTIGE in H$\alpha$ if they contain ionised gas. All these sources will be searched for in the deep 
NB H$\alpha$ images. Thanks to the sensitivity of VESTIGE we expect to detect in H$\alpha$ a large fraction 
of HVC and LVC discovered by the HI GALFA and ALFALFA surveys, and thus significantly increase the number of objects with 
both atomic and ionised gas data, now limited to a handful of sources (Reynolds 1987; Haffner et al. 2001; Tufte et al. 2002). 
The typical peak emissivity of these features is $\simeq$ 0.1-0.5 Rayleigh (1 R = 5.66 $\times$ 10$^{-18}$ erg s$^{-1}$ cm$^{-2}$ arcsec$^{-2}$;
Reynolds 1987; Haffner et al. 2001; Tufte et al. 1998, 2002; Putman et al. 2003), and their extension is of several arcminutes (Saul et al. 2012), thus easily detectable 
after spatial smoothing of the VESTIGE data. The H$\alpha$ data, combined with HI and X-rays, will be used to constrain the physical
properties of the clouds, study their Galactic or extragalactic origin, and quantify the escape radiation from the Galactic plane (Bland-Hawthorn \& Maloney 1999; 
Putman et al. 2003, 2012) by comparing models to observations (Heitsch \& Putman 2009; Binney et al. 2009; Wood et al. 2010; Kwak \& Shelton 2010;
Shelton et al. 2012).

\subsection{Background science}

\subsubsection{High redshift emission line galaxies}

The H$\alpha$ filter combined with broad-band $r$ or deeper NGVS images will allow us to identify strong Ly$\alpha$, [OII] and [OIII] 
emission-line galaxies at different redshifts ($z$=4.4,0.8,0.3), providing a unique sample of star-forming objects over $\sim$ 104 deg$^2$. 
Indeed, the point-source sensitivity of VESTIGE ($f(H\alpha$) = 4 $\times$ 10$^{-17}$ erg sec$^{-1}$ cm $^{-2}$, 5$\sigma$ detection limit) 
is comparable to the typical Ly$\alpha$ luminosity of $z=$4.4 galaxies as derived from 
their Ly$\alpha$ luminosity function ($L^*$ $\sim$ 2-4 $\times$ 10$^{-17}$ erg sec$^{-1}$ cm $^{-2}$; 
Ouchi et al. 2008; Cassata et al. 2011). It is also $\sim$ 2$\sigma$ and $\sim$ 10$\sigma$ deeper than the [OII] and [OIII] 
$L^*$ luminosities at $z$ = 0.8 and $z$ =0.3 
(Ly et al. 2007; Ciardullo et al. 2013). Using model predictions for the Ly$\alpha$ line (Garel et al. 2015, 2016), or the luminosity functions and 
the number counts of [OII] and [OIII] 
sources derived with similar surveys of small fields (Ly et al. 2007; Hippelein et al. 2003), we estimate that $\sim$ 5 $\times$ 10$^4$ Ly$\alpha$, 
$\sim$ 10$^5$ [OII], and 3 $\times$ 10$^4$ 
[OIII] emitters will be detected over $\sim$ 104 deg$^2$. VESTIGE will provide an unprecedented database for characterising the bright end of the 
high-redshift emission line galaxies luminosity function, insensitive to cosmic variance. The data will thus be highly complementary to those obtained 
in deep targeted imaging and spectroscopic surveys, which are generally limited to $\lesssim$ 1 deg$^2$ (and optimised for studying the 
luminosity function 
at its faint end). At the depth of VESTIGE, the majority of the detected sources will be background objects. The continuum-subtracted image, 
however, will contain only local H$\alpha$ or background line emitters. High-redshift [OII] and [OIII] sources will be identified using photometric 
redshifts available from the NGVS (Raichoor et al. 2014), while Ly$\alpha$ candidates using standard 
(H$\alpha$-$r$) vs. H$\alpha$ colour magnitude 
relations (Ouchi et al. 2008) as shown in Fig. \ref{pointsource}, combined with other relations based on deeper $g$ (25.9 mag) and $i$ (25.1 mag; 10$\sigma$) NGVS images. 
Multi-slit wide-field 
spectroscopic follow-up will be carried out on a subset of the candidates 
to quantify the statistically accuracy in the photometric identification of the different line emitting sources.

\subsubsection{QSOs}

\begin{figure*}
   \centering
   \includegraphics[width=9cm]{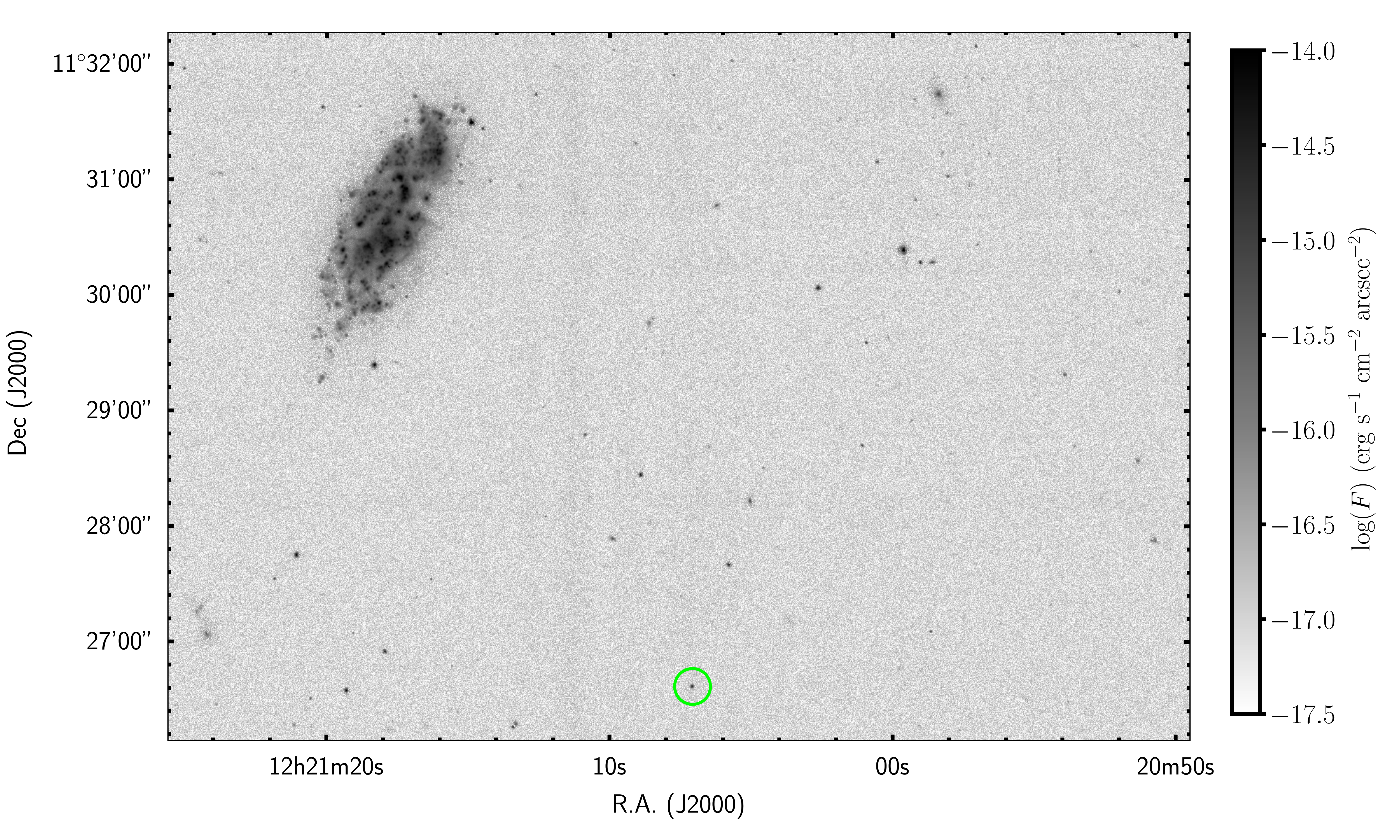}
   \includegraphics[width=9cm]{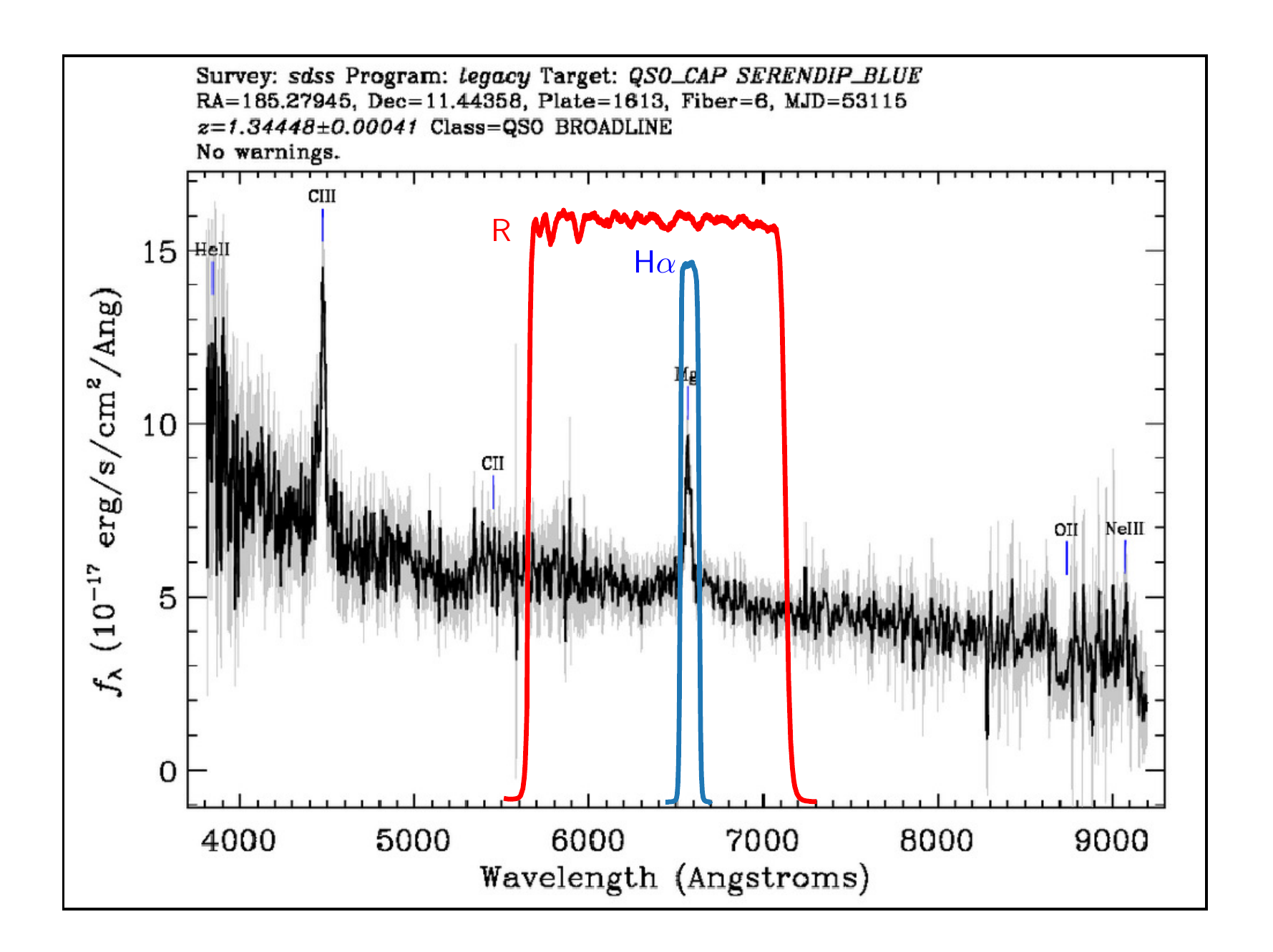}
   \caption{Left: the continuum-subtracted H$\alpha$ image of NGC 4294. The point source marked with a green circle indicates
   SDSS J122107.07+112636.8, a QSO at $z$=1.34. Right: the SDSS spectrum of this source shows a strong MgII ($\lambda$ 2799 \AA) emission line
   in the NB filter centred on the rest-frame H$\alpha$.}
   \label{QSO}%
   \end{figure*}

VESTIGE will be able to detect background quasars whenever one of their prominent emission lines falls within the narrow band filter. 
The most important emission lines in the spectrum of a quasar are [OIII] ($\lambda$ 5007 \AA; $z$=0.3), H$\beta$ ($\lambda$ 4861 \AA; $z$=0.35),
MgII ($\lambda$ 2799 \AA; $z$=1.3), CIII ($\lambda$ 1908 \AA; $z$=2.4), CIV ($\lambda$ 1549 \AA; $z$=3.2, and Ly$\alpha$ ($\lambda$ 1216 \AA; $z$=4.4)
(Vanden Berk et al. 2001). A typical example is the object SDSS J122107.07+112636.8, identified by SDSS as a background QSO at $z$=1.344, detected
as a strong point source in the proximity of the galaxies NGC 4294 in the continuum-subtracted H$\alpha$ image obtained during pilot observations of the VESTIGE
survey (Fig. \ref{QSO}). The QSO has been detected thanks to a prominent MgII line emission.

\subsubsection{Photometric redshifts}

The $r$-band frames that will be gathered to subtract the contribution of the stellar continuum from the NB emission will be combined to
those already available from the NGVS ($ugiz$) to increase the accuracy of the photometric redshift of background sources. At a limiting magnitude of $r$ $\sim$ 25.3
ABmag (5$\sigma$ detection limit for point sources), the addition of the $r$-band will increase the accuracy of photometric redshifts by $\sim$ 30\%~ in the faintest
objects and reduce the number of catastrophic outliers by more than a factor of two (Raichoor et al. 2014). The accuracy of the photometric redshift can be improved 
by the inclusion of a NB filter (COMBO-17, Wolf et al. 2003), or of other UV and near-IR photometric bands (COSMOS, Ilbert et al. 2009; NMBS, Whitaker et al. 2011). 
The NB filter will be crucial for the identification of overdense regions at those redshifts where it will detect the brightest emission lines 
of the spectrum ([OIII], H$\beta$, [OII], Ly$\alpha$). 
Photometric redshifts are fundamental for the identification of cluster galaxies using a red sequence technique, particularly in the range 0.3 $<$ $z$ $<$ 0.8 where the $r$-band includes
the 4000 \AA~ break (Raichoor et al. 2014). We expect to detect $\sim$ 4000 clusters in the 0.1 $<$ $z$ $<$ 1.0 redshift range over the 104 deg$^2$
covered by the NGVS survey (Olsen et al. 2007; Licitra et al. 2016), a number comparable to that discovered by the CFHTLS-Wide survey (Durret et al. 2011).

\section{Synergy with other surveys and follow-up observations}

VESTIGE is largely a self-contained project, but will capitalise on the extensive set of 
multifrequency data available for the cluster (much of it obtained by members of the VESTIGE team).\\

\noindent
\textit{X-ray}: X-ray data are necessary to trace the distribution of the hot ICM (ROSAT - B\"ohringer et al. 1994; ASCA - Kikuchi et al. 2000) or
to quantify the physical properties of the hot gas phase in the stripped material of perturbed galaxies. 
There are 131 XMM observations and 365 \textit{Chandra} observations between 
0.5 deg and 8 deg from M87 in the archives. These XMM and \textit{Chandra} observations also include many programs dedicated to Virgo spirals, such as the 2016 \textit{Chandra} 
large project ``Spiral Galaxies for the Virgo Cluster'' led by Dr. Roberto Soria (559 ks on 52 galaxies). 

\noindent
\textit{UV, visible, and near-infrared}: UV, visible, and near-infrared data are necessary to characterise the stripping process, identify 
possible ionising sources in the stripped material, and detect the optical and near-infrared counterparts of H$\alpha$ point sources. 
The whole Virgo cluster region has been mapped in the UV by GALEX (GUViCS - Boselli et al. 2011, Voyer et al. 2014) and in the visible by the SDSS (York et al. 2000),
NGVS (Ferrarese et al. 2012), and Pan-STARRS (Magnier et al. 2013). The cluster has been also a favorite target of HST (e.g., ACSVCS, Cot\'e et al. 2004).
Near-infrared imaging data in the $K_S$-band over the 4 deg$^2$ centred on M87, and 
in the $J$- and $K_S$-band over $\sim$ 16 deg$^2$ in the region between M87 and M49 have been taken during the NGVS-IR project (Munoz et al. 2014).

\noindent
\textit{Mid- and far-infrared}: Infrared data are essential for correcting H$\alpha$ data for dust attenuation, for identifying the emission 
of the Galactic cirrus, and for a complete characterisation of the spectral energy distribution of local and background sources.
Mid- (4-22 $\mu$m) and far-IR (70-500 $\mu$m) data are available from WISE (Wright et al. 2010) and 
from the \textit{Herschel} blind survey HeViCS (Davies et al. 2010; Auld et al. 2013), while pointed observations of the brightest galaxies from the \textit{Herschel}
Reference Survey (HRS; Boselli et al. 2010, Ciesla et al. 2012, Cortese et al. 2014). ISO and \textit{Spitzer} observations are available for a large fraction of the
brightest galaxies (Boselli et al. 2003, 2014a, Bendo et al. 2012, Ciesla et al. 2014), while in the sub-millimetre from the \textit{Planck} space mission
(Planck collaboration 2014; Baes et al. 2014). 

\noindent
\textit{Radio millimetre and centimetre}: HI and CO data are crucial for quantifying the physical properties of the cold gas phase over the disc of galaxies and within
the stripped material. Radio continuum observations, sensitive to the energy loss of relativistic electrons spinning in weak magnetic fields,
will provide a further tracer of ongoing perturbations as often done through the identification of head-tail radio
sources (e.g. Ulrich 1978; Sarazin 1986).
The Virgo cluster region has been fully mapped in the HI line by ALFALFA (Giovanelli et al. 2005; Haynes et al. 2011) with Arecibo, while pointed high-resolution 
observations are available from the VLA VIVA survey (Chung et al. 2009a). The cluster will also be fully mapped by the Wallaby survey 
($rms$ $\sim$ 1.6 mJy at 30 arcsec and 4 km s$^{-1}$ resolution; Koribalski 2012). CO data are available for the brightest targets (Kenney \& Young 1988; 
Young et al. 1995; Boselli et al. 1995, 2002, 2014b; Helfer et al. 2003; Chung et al. 2009b). 
Radio continuum observations are available from the NVSS (Condon et al. 1998) and FIRST surveys (Becker et al. 1995; see Gavazzi \& Boselli 1999)
and will be soon available from the ASKAP EMU survey for the whole Virgo cluster region at a sensitivity of rms $\sim$ 10 $\mu$Jy/beam at 1.3 GHz with an angular
resolution of 10 arcsec (Norris et al. 2011).

\noindent
\textit{Spectroscopy}: Multi-slit wide-field spectroscopy is necessary for the identification and the characterisation of point sources,
while IFU spectroscopy is required for the study of the physical of the perturbed galaxies and of the stripped gas. IFU data are available for
a small fraction of the early-type galaxies from Guerou et al. (2015) or from the ATLAS-3D survey (Cappellari et al. 2011), 
and for several late-type systems from Chemin et al. (2006).
High resolution long-slit spectra are also available for a dozen of dwarves (Toloba et al. 2011, 2014).

\section{Summary}

The Virgo Environmental Survey Tracing Ionised Gas Emission (VESTIGE) is a blind NB H$\alpha$
imaging survey of the Virgo cluster region up to its virial radius. The survey, started in 2017 and planned to run for three years, is 
carried out with MegaCam at the Canada-France-Hawaii Telescope. VESTIGE reaches a sensitivity of 
$f(H\alpha)$ $\sim$ 4 $\times$ 10$^{-17}$ erg sec$^{-1}$ cm$^{-2}$ (5$\sigma$ detection limit) for point sources and
$\Sigma (H\alpha)$ $\sim$ 2 $\times$ 10$^{-18}$ erg sec$^{-1}$ cm$^{-2}$ arcsec$^{-2}$ (1$\sigma$ detection limit at 3 arcsec resolution) 
for extended sources, and will be the deepest and largest blind NB survey of a nearby cluster. 
The observations carried out so far of the centre of the cluster show that, at this sensitivity,
VESTIGE is able to detect extended filaments of ionised gas produced by the interaction of galaxies with the surrounding environment. 
This survey has been designed to study the effects of the environment on galaxy evolution and
will provide, for years to come, an ideal reference for comparison with cosmological models. 
As designed, VESTIGE will also be used to study
the fate of the stripped gas in cluster objects, the star formation process in nearby galaxies of
different type and stellar mass, the determination of the H$\alpha$ luminosity function and of the
H$\alpha$ scaling relations down to $\sim$ 10$^6$ M$_{\odot}$ stellar mass objects, and the reconstruction 
of the dynamical structure of the Virgo cluster. Thanks to its sensitivity and large sky coverage, 
VESTIGE will also be used to study the HII luminosity function on hundreds of galaxies, the diffuse H$\alpha$ emission of 
the Milky Way at high Galactic latitude, and the properties of emission line galaxies at high redshift.
The legacy value of VESTIGE is thus very high and the survey is virtually guaranteed to provide a reference for years to come. 

\begin{acknowledgements}

We thank the anonymous referee for useful comments on the manuscript.
We thank Jeffrey Chan for useful advices on the determination of the survey depth.
We acknowledge financial support from "Programme National de Cosmologie and Galaxies" (PNCG) funded by CNRS/INSU-IN2P3-INP, CEA and CNES, France,
and from "Projet International de Coop\'eration Scientifique" (PICS) with Canada funded by the CNRS, France.
This research has made use of the NASA/IPAC Extragalactic Database (NED) 
which is operated by the Jet Propulsion Laboratory, California Institute of 
Technology, under contract with the National Aeronautics and Space Administration
and of the GOLDMine database (http://goldmine.mib.infn.it/) (Gavazzi et al. 2003b).
MB was supported by MINEDUC-UA projects, code ANT 1655 and ANT 1656.
AL work was supported by the Sino-French LIA-Origin joint program.
MS acknowledges support from the NSF grant 1714764 and the Chandra Award GO6-17111X.
MF acknowledges support by the science and technology facilities council [grant number ST/P000541/1].
EWP acknowledges support from the National Natural Science Foundation of China through Grant No. 11573002.
KS acknowledges support from the Natural Sciences and Engineering Research Council of Canada (NSERC).

\end{acknowledgements}

\end{document}